\documentclass[twocolumn, a4size, 10pt]{IEEEtran}

\IEEEoverridecommandlockouts

\makeatletter
\def\markboth#1#2{\def\leftmark{\@IEEEcompsoconly{\sffamily}\MakeUppercase{\protect#1}}%
\def\rightmark{\@IEEEcompsoconly{\sffamily}\MakeUppercase{\protect#2}}}
\makeatother

\usepackage[T1]{fontenc}

\usepackage{kantlipsum}
\usepackage{setspace}

\usepackage[english]{babel}
\usepackage{xcolor}
\usepackage{lipsum}
\usepackage{caption}
\usepackage{cite}
\usepackage[pdftex]{graphicx}
\usepackage{subcaption}
\usepackage{amsmath}
\usepackage{booktabs}
\usepackage{colortbl}
\usepackage{amsfonts}
\usepackage{amssymb}
\usepackage{amsthm}
\usepackage{array}
\usepackage{ragged2e}
\usepackage{dblfloatfix} 

\usepackage{verbatim}
\usepackage{listings}
\usepackage{algorithm}
\usepackage{algpseudocode}
\usepackage{url}
\usepackage{enumerate}
\usepackage{multirow}

\definecolor{LightBlue}{rgb}{0.5,0.5,1}
\definecolor{LightRed}{rgb}{1,0.5,0.5}
\definecolor{LightYellow}{rgb}{1,0.85,0}

\usepackage{epsfig}
\usepackage{epstopdf}
\usepackage{multicol}
\usepackage[font=footnotesize]{caption}

\usepackage{algorithm}

\makeatletter
\def\BState{\State\hskip-\ALG@thistlm}
\makeatother

\renewcommand{\arraystretch}{2}

\newcommand{\E}{\mathbf{E}}

\newcommand{\bi}{\begin{itemize}}
\newcommand{\ei}{\end{itemize}}
\newcommand{\be}{\begin{align}}
\newcommand{\ee}{\end{align}}

\addto\captionsenglish{}

\addto\captionsenglish{}

\newtheorem{proposition}{Proposition}
\newtheorem{remark}{Remark}

\def\beq{\begin{align}}
\def\eeq{\end{align}}
\def\beqa{\begin{eqnarray}}
\def\eeqa{\end{eqnarray}}
\def\beqan{\begin{eqnarray*}}
\def\eeqan{\end{eqnarray*}}

\def\SINR{\mathsf{SINR}}
\def\SIR{\mathsf{SIR}}

\def\({\left(}
\def\){\right)}
\def\[{\left[}
\def\]{\right]}
\newcommand{\nn}{\nonumber}

\def\tx{\text{TX}}
\def\rx{\text{RX}}

\def\3G{\mathsf{3GPP}} 

\def\ISO{\mathsf{ISO}}

\def\C{C_{\SINR}}
\def\hC{\hat{C}_{\SINR}}

\title{\fontsize{19}{24}\selectfont Stochastic Geometric Coverage Analysis in mmWave Cellular Networks with Realistic Channel and Antenna Radiation Models}

\author{ Mattia Rebato, \emph{Student Member, IEEE}, Jihong Park, \emph{Member, IEEE}, Petar Popovski, \emph{Fellow, IEEE}, Elisabeth De Carvalho, \emph{Senior Member, IEEE}, Michele Zorzi, \emph{Fellow, IEEE}
\thanks{M. Rebato and M. Zorzi are with the Department of Information Engineering, University of Padova, 35122 Padova, Italy (email: $\{$rebatoma, zorzi$\}$@dei.unipd.it).}
\thanks{J. Park is with the Centre for Wireless Communications, University of Oulu, 90014 Oulu, Finland (email: jihong.park@oulu.fi).}
\thanks{P. Popovski and E. de Carvalho are with the Department of Electronic Systems, Aalborg University, 9100 Aalborg, Denmark (email: $\{$petarp,edc$\}$@es.aau.dk).}
\thanks{A preliminary version of this paper was presented at the IEEE GLOBECOM conference, December 2017~\cite{rebato17}.
This work has benefited from comments and suggestions by M. Haenggi.
The work of P. Popovski has been partially supported by the Danish Ministry of Higher Education and Science (EliteForsk Award, Grant Nr. 5137-00073B).
The work of E. de Carvalho has been partially supported by the Danish Council for Independent Research (DFF701700271).
The work of M. Zorzi has been partially supported by the Villum Foundation, Denmark.
}
}

\begin{document}
\maketitle
\begin{abstract} 

Millimeter-wave (mmWave) bands will play an important role in 5G wireless systems.
The system performance can be assessed by using models from stochastic geometry that cater for the directivity in the desired signal transmissions as well as the interference, and by calculating the signal-to-interference-plus-noise ratio ($\SINR$) coverage.
Nonetheless, the accuracy of the existing coverage expressions derived through stochastic geometry may be questioned, as it is not clear whether they capture the impact of the detailed mmWave channel and antenna features.
In this study, we propose an $\SINR$ coverage analysis framework that includes realistic channel model and antenna element radiation patterns.
We introduce and estimate two parameters, \emph{aligned gain} and \emph{misaligned gain}, associated with the desired signal beam and the interfering signal beam, respectively. 
The distributions of these gains are used to determine the distribution of the $\SINR$ which is compared with the corresponding $\SINR$ coverage calculated via system-level simulations.
The results show that both aligned and misaligned gains can be modeled as \emph{exponential-logarithmically} distributed random variables with the highest accuracy, and can further be approximated as \emph{exponentially} distributed random variables with reasonable accuracy.
These approximations can be used as a tool to evaluate the system-level performance of various 5G connectivity scenarios in the mmWave band.
\end{abstract}
\smallskip
\begin{IEEEkeywords}
Millimeter-wave, channel model, antenna radiation pattern, large-scale cellular networks, stochastic geometry. 
\end{IEEEkeywords}
\vspace{-0.5cm}

\section{Introduction}
\label{introduction}
Millimeter-wave (mmWave) frequencies can provide $20$-$100$ times larger bandwidth than current cellular systems.
To enjoy this benefit in 5G cellular systems, the significant distance attenuation of the desired mmWave signals needs to be compensated by means of sharpened transmit/receive beams~\cite{roh2014,mmwave3gpp}.
The directionality of mmWave transmissions can induce intermittent yet strong interference to the neighboring receivers.
The sharpening of the directional beams reduces the probability of interference from the mainlobe, while increasing the signal strength within the mainlobe. This has a significant impact on the statistics of the signal-to-interference-plus-noise ratio ($\SINR$) across the network.

In this paper, we incorporate the experimental models for mmWave channels and antenna radiations into the tools of stochastic geometry.
This results in a sufficiently realistic framework for system-level analysis of mmWave systems.
Fig.~\ref{globalview} illustrates the framework, which can be seen as a semi-heuristic, as it bridges the gap between a very theoretical study at a large scale (stochastic-geometric analysis), and practical measurements at a small scale.
The novelty of our work compared to the existing works on mmWave $\SINR$ coverage analysis is summarized in the following subsections.

\subsection{Background and Related Works}
The $\SINR$ coverage of a mmWave cellular network has been investigated in~\cite{wang1,bai15,Andrews:17,direnzo2015,park2016,li16,Gupta2016,Kim:18} using stochastic geometry, a mathematical tool able to capture the random interference behavior in a large-scale network.
Compared to traditional cellular systems using sub-$6$~GHz frequencies, the major technical difficulty of mmWave $\SINR$ coverage analysis comes from incorporating their unique channel propagation and antenna radiation characteristics in a tractable way, as detailed next.

\subsubsection{Channel gain model}
mmWave signals are vulnerable to physical blockages, which can lead to significant distance attenuation under non-line-of-sight (NLoS) channel conditions as opposed to under line-of-sight (LoS) conditions.
This is incorporated in the mmWave path loss models by using different path loss exponents for LoS and NLoS conditions.
Besides this large-scale channel gain, there exists a small-scale fading due to reflections and occlusions by human bodies.
In order to capture this, while maximizing the mathematical tractability, one can introduce an \emph{exponentially} distributed gain as done in~\cite{park2016,li16,Gupta2016,Kim:18}.
This implies assuming Rayleigh fading, which is not always realistic, particularly when modeling the sparse scattering characteristics of mmWave signals~\cite{park2016}.

At the cost of making analytical tractability more difficult, several works have detoured this problem by considering generalized small-scale channel gains that follow a \emph{gamma} distribution (i.e., Nakagami-$m$ fading)~\cite{bai15,Andrews:17,wang1} or a \emph{log-normal} distribution~\cite{direnzo2015}.
Nevertheless, such generic fading models have not been compared with real mmWave channel measurements, and may therefore either overestimate or underestimate the actual channel behaviors.

\begin{figure*}[h!]
\centering
\begin{subfigure}[b]{0.48\textwidth}
\centering
\includegraphics[width=1\textwidth]{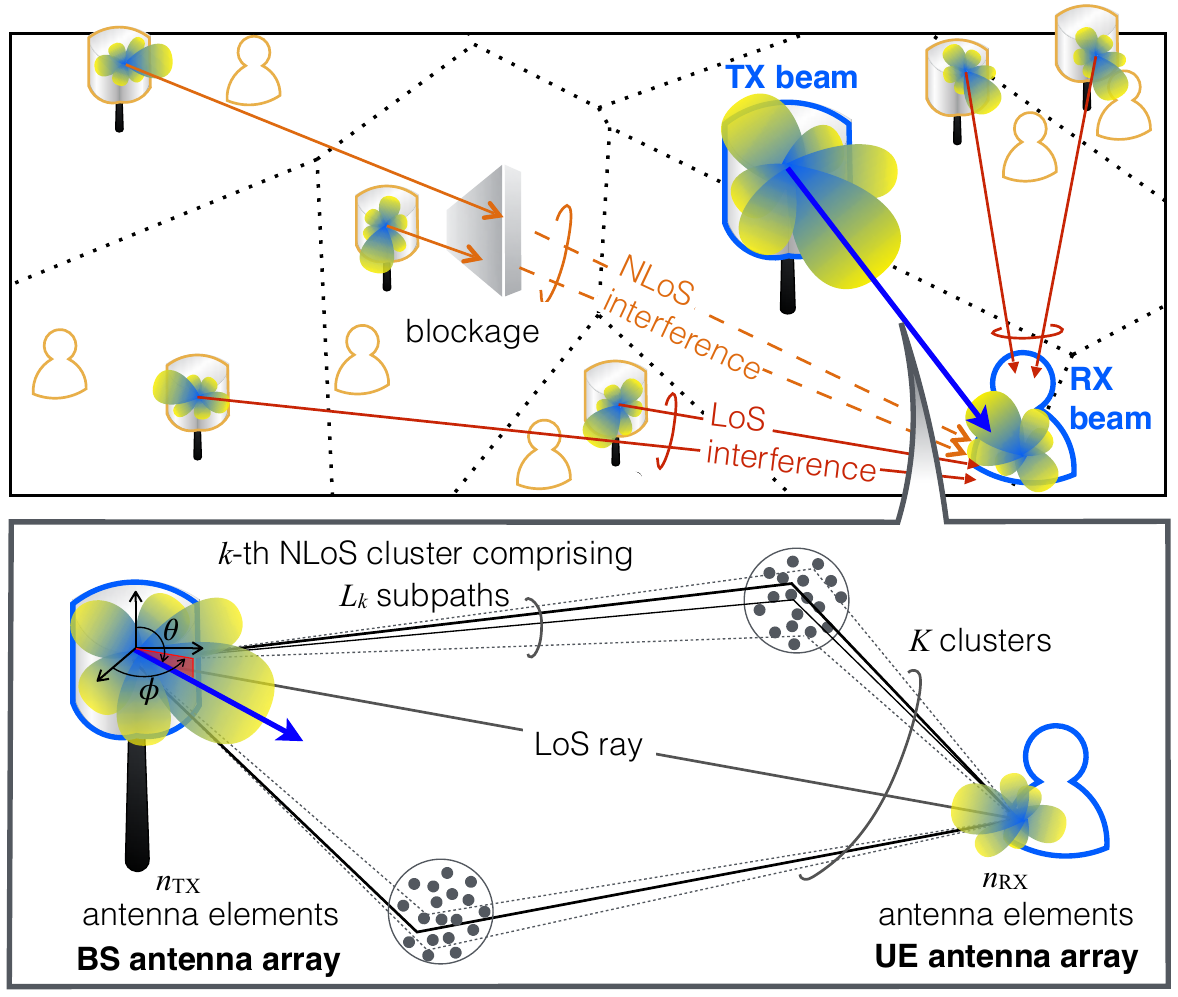}
\caption{With the $\ISO$ element pattern~\cite{rebato17}.}
\label{ISOnetwork}
\end{subfigure}
\hfill
\begin{subfigure}[b]{0.48\textwidth}
\centering
\includegraphics[width=1\textwidth]{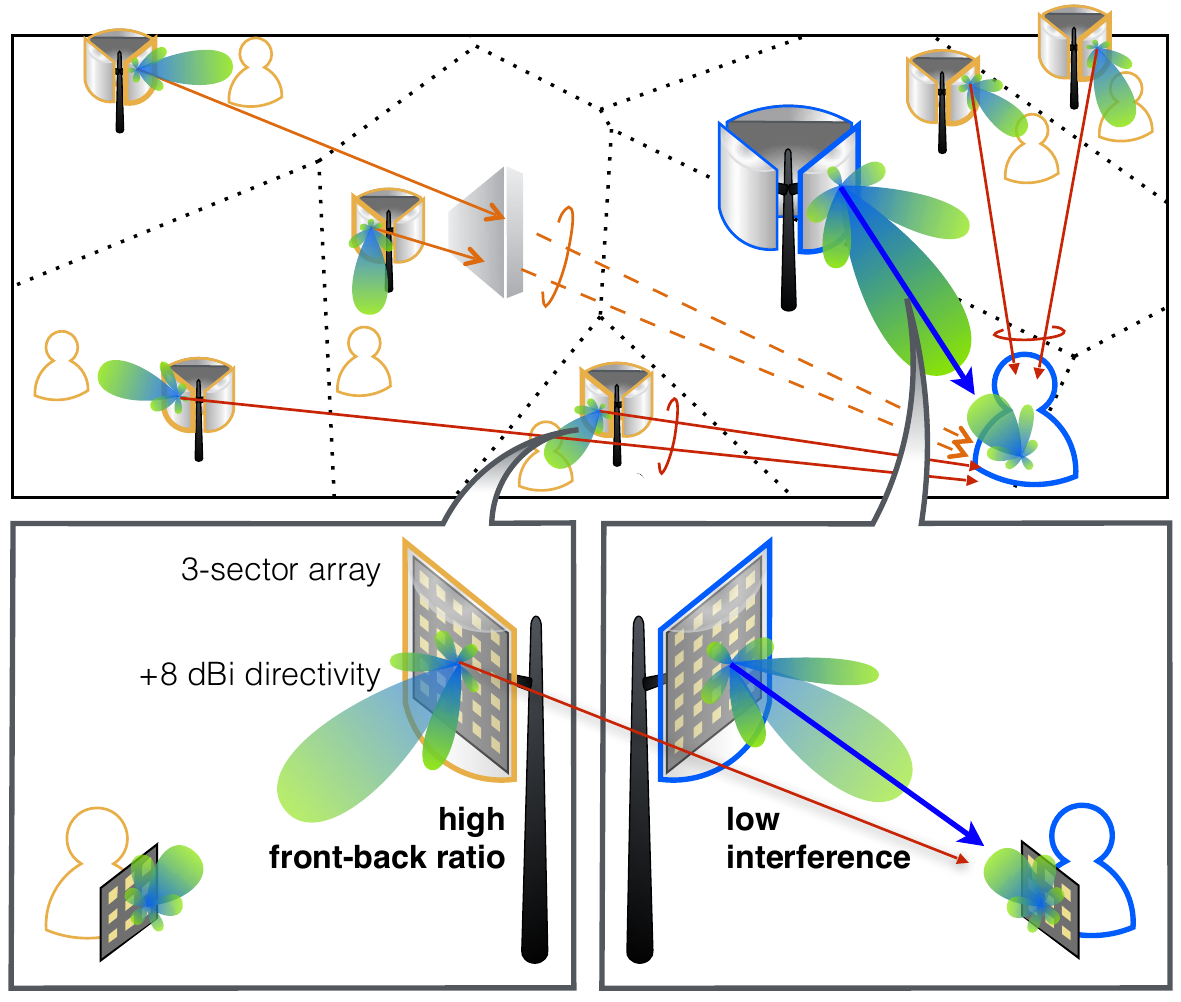}
\caption{With the $\3G$ element pattern~\cite{mmWave_3gpp_channel}.}
\label{3GPPnetwork}
\end{subfigure}
\caption{An illustration of our mmWave network model (top) and the channel model of each link with the transmitter/receiver antenna radiation model (bottom):\\ (a) With the $\ISO$ element pattern, antenna gain parameters come from our previous work~\cite{rebato17}; (b) With the $\3G$ element pattern, antenna gain parameters follow from the $\3G$ specifications~\cite{mmWave_3gpp_channel}. For both element radiation patterns, channel parameters are obtained from a measurement-based mmWave channel model provided by the NYU Wireless~Group~\cite{akdeniz14}.}
\label{globalview}
\end{figure*}

\subsubsection{Antenna gain model}
Both base stations (BSs) and user equipments (UEs) in 5G mmWave systems are envisaged to employ planar antenna arrays that enable directional transmissions and receptions.
A planar antenna array comprises a set of patch antenna elements placed in a two-dimensional plane.
The radiation pattern of each single antenna element is either isotropic or directional, which are hereafter denoted as $\ISO$ and $\3G$ element patterns, respectively.
By superimposing the radiation of all the antenna elements, a planar antenna array is able to enhance its radiation in a target direction while suppressing the radiation in other directions. 

The $\3G$ element pattern is incorporated in the antenna gain model provided by the 3GPP~\cite{mmWave_3gpp_channel}.
Compared to the $\ISO$ element pattern, the directional antenna elements in the $\3G$ element pattern enable element-wise beam steering, thereby yielding higher mainlobe and lower sidelobe gains, i.e., increased front-back ratio\footnotemark, as visualized in Fig.~\ref{3GPPnetwork}.
Such benefit diminishes as the beam steering direction becomes closer to the plane of the antenna array.
In order to solve this problem, the 3GPP suggests to equip each BS with $3$-sectored antenna arrays~\cite{mmWave_3gpp_channel}, thus restricting the beam steering angle to $\pm 60^\circ$.

\footnotetext{The front-back ratio is the difference expressed in decibels between the gain of the mainlobe and the second maximum gain. This ratio increases with the number of antenna elements~\cite{tse_book,rebato18}.}

The said radiation characteristics and antenna structure of the $\3G$ element pattern complicate the antenna gain analysis.
For this reason, most of the existing approaches based on stochastic geometry~\cite{rebato17,bai15,direnzo2015,Andrews:17,park2016,li16,Gupta2016,Kim:18,Andrews:17bis} still resort to the $\ISO$ element pattern.
This underestimates the front-back ratio of the actual cellular system, degrading the accuracy in the mmWave $\SINR$ coverage analysis.
Furthermore, the antenna gains are commonly approximated by using two constants obtained from the maximum and the second maximum lobe gains~\cite{bai15,direnzo2015,Gupta2016,Andrews:17,park2016,li16,Kim:18}.
It is unclear whether such an approximation is still applicable for the mmWave $\SINR$ coverage analysis with realistic radiation patterns.
By approximating the original system model with a simplified one, whose performance is determined by a mathematically convenient intensity measure, tractable yet accurate integral expressions for computing area spectral efficiency and potential throughput are provided in~\cite{diRenzo2016}.
The considered system model accounts for many practical aspects which are typically neglected, e.g., LoS and NLoS propagation, antenna radiation patterns, traffic load, practical cell associations, and general fading channels.
However, a measurement-based channel characterization is missing. 

Recently, a few studies~\cite{Haenggi17} and~\cite{Haenggi18} incorporate the impact of directional antenna elements on the stochastic geometric $\SINR$ coverage analysis, by approximating the element radiation pattern as a cosine-shaped curve under a one-dimensional linear array structure.
Compared to these works, we consider two-dimensional planar arrays, and approximate the combined array-and-channel gain as a single term, as detailed in the following subsection.

\subsubsection{Aligned/misaligned gain model}
In order to solve the aforementioned issue brought by inaccurate channel gains, one can use measurement-based channel gain models, such as the models provided by the New York University (NYU) Wireless Group~\cite{akdeniz14}, which are operating at 28~GHz as described in~\cite{akdeniz14,samimi15,mezzavilla15,ford16}.
However, the NYU channel gain model requires a large number of parameters, and is thus applicable only to system-level simulators with high complexity, as done in our previous study~\cite{rebato16}.

In our preliminary work~\cite{rebato17}, we simplified the NYU channel gain model via the following procedure so as to allow stochastic geometric $\SINR$ coverage analysis.
\begin{enumerate}[(i)]
	\item We separated the path loss gains from the small-scale fading, and treated them independently in a stochastic geometric framework.
	The fading term can be considered as representative of propagation effects when the user moves locally, and is independent of the link distance.

	\item For each downlink communication link, we combined the channel gain and the antenna gain into an aggregate gain.
	The aggregate gain is defined for the desired communication link as \emph{aligned gain} and for an interfering link as \emph{misaligned gain}, respectively.

	\item We applied a curve fitting method to derive the distributions of the aligned/misaligned gains.

	\item Finally, we derived the distribution of a reference user's $\SINR$, which is a function of path loss gains and aligned/misaligned gains, by applying a stochastic geometric technique to the path loss gains and then by exploiting the aligned/misaligned gain distributions. 
\end{enumerate}

The limitation of our previous work~\cite{rebato17} is its use of the $\ISO$ element pattern in step (ii).
This results in excessive sidelobe gains, particularly including backward propagation, which are unrealistic.
To fix this problem, in this study we also apply the $\3G$ element pattern to the aforementioned aligned/misaligned gain model, thereby yielding a tractable mmWave coverage expression that ensures high accuracy, comparable to the results obtained from a system-level simulator.
Moreover, instead of signal-to-interference ratio ($\SIR$) as considered in~\cite{rebato17}, we focus on the $\SINR$ evaluation by incorporating also the impact of the noise power.

A recent work~\cite{Andrews:17bis} is relevant to this study. While neglecting interference, it firstly considers a simplified keyhole channel, and then introduces a correction factor.
The aggregate channel gain thereby approximates the channel gain under the mmWave channel model provided by the $\3G$~\cite{3gppChannel:17}.
Compared to this, using the NYU channel model~\cite{akdeniz14}, we additionally consider a realistic antenna radiation pattern provided by the $\3G$~\cite{mmWave_3gpp_channel}.
In addition, we explicitly provide the $\SINR$ coverage probability expression using these realistic channel and antenna models, as well as its simplified expression.

\subsection{Contributions and Organization}
The contributions of this paper are summarized below.
\begin{itemize}
	\item Accurate distributions of aligned and misaligned gains are provided (see \textbf{Remarks 1}-\textbf{4}), which reflect the NYU mmWave channel model~\cite{akdeniz14} and the 3GPP mmWave antenna radiation model~\cite{mmWave_3gpp_channel}.
	
	\item Considering the $\ISO$ element pattern, following from our preliminary study~\cite{rebato17}, the aligned gain is shown to follow an \emph{exponential} distribution, despite the scarce multipath in mmWave channels (\textbf{Remark~1}).
	On the other hand, we show that the misaligned gain can be approximated with a \emph{log-logistic} distribution (\textbf{Remark~3}) having a heavier tail than the exponential distribution, which can be lower and upper bounded by a \emph{Burr} distribution and a \emph{log-normal} distribution, respectively.
	
	\item In contrast, for the 3GPP element pattern, we show that both aligned and misaligned gains independently follow an \emph{exponential-logarithmic} distribution (\textbf{Remarks~2} and \textbf{4}), which has a lighter tail compared to the exponential distribution.

	\item Applying these aligned and misaligned gain distributions, the downlink mmWave $\SINR$ coverage probabilities with the $\ISO$ and $\3G$ element patterns are derived using stochastic geometry (\textbf{Propositions 1} and~\textbf{2}).
\end{itemize}

In spite of the exponential-logarithmical distribution of the aligned/misaligned gains of the $\3G$ element pattern, it is still possible, in the $\SINR$ calculation, to approximate both gains independently using \emph{exponential} random variables with proper mean value adjustment (\textbf{Remark~5} and Fig.~\ref{global_comparison}), yielding a further simplified (though slightly less accurate) $\SINR$ coverage probability expression (\textbf{Proposition 3}).
The feasibility of the exponential approximation under the $\3G$ element pattern comes from the identical tail behaviors of both aligned/misaligned gains, that cancel each other out during the $\SINR$ calculation.
Following the same reasoning, this approach provides a similar approximation under the $\ISO$ element pattern that leads to the different tail behaviors of both the aligned/misaligned gains due to the low front-back ratio obtained with isotropic elements (see Fig.~\ref{global_comparison} in Sect.\ref{numerical_results}).

The remainder of this paper is organized as follows.
Section~\ref{system_model} describes the channel model and antenna radiation patterns.
Section~\ref{Gain_distribution} proposes the approximated distributions of aligned and misaligned gains. Section~\ref{sinr_coverage} derives the $\SINR$ coverage probability. Section~\ref{numerical_results} validates the proposed approximations and the resulting $\SINR$ coverage probabilities by simulation, followed by our conclusion in Section~\ref{conclusion_and_future_works}.

\section{System Model}
\label{system_model}
In this study, we consider a downlink mmWave cellular network where both BSs and UEs are independently and randomly distributed in a two-dimensional Euclidean plane.
Each UE associates with the BS that provides the maximum average received power, i.e., minimum path loss association.
The UE density is assumed to be sufficiently large such that each BS has at least one associated UE.
Multiple UEs can be associated with a single BS, while the BS serves only a single UE per unit time slot according to a uniformly random scheduler, as assumed in \cite{park2016,rebato17,Kim:18} under stochastic geometric settings.
Out of these serving users in the network, we hereafter focus on a reference user that is located in the origin of the area considered, and is denoted as the typical UE.
This typical UE's $\SINR$ is affected by the antenna array radiation patterns and channel gains, as described in the following subsections.

\subsection{Antenna Gain}
\label{antenna_gain}
Each antenna array at both BS and UE sides contributes to the received signal power, according to the radiation patterns of the antenna elements that comprise the antenna array. The amount is affected also by the vertical angle $\theta$, horizontal angle $\phi$, and polarization slant angle $\zeta$, as described next.

\subsubsection{Element radiation pattern}
For each antenna element in an antenna array, we consider two different radiation patterns: isotropic radiation and the radiation provided by the $\3G$~\cite{antenna_3gpp}. The element radiation pattern $A_E^{(z)}(\theta,\phi)$ (dB) for superscript $z\in\{\ISO,\3G\}$ specifies how much power is radiated from each antenna element towards the direction $(\theta,\phi)$.

Following our preliminary study~\cite{rebato17}, with the $\ISO$ element pattern, each antenna element radiates signals isotropically with equal transmission power. Hence, for all $\theta\in[0,180^\circ]$ and $\phi\in[-180^\circ,180^\circ]$, the $\ISO$ element radiation pattern is given~as
\begin{align}
A_E^{(\ISO)}(\theta,\phi)=0~\text{dB}.
\end{align}

\begin{figure*}[h!]
\centering
\begin{subfigure}[b]{0.45\textwidth}
\centering
\includegraphics[width=1\textwidth]{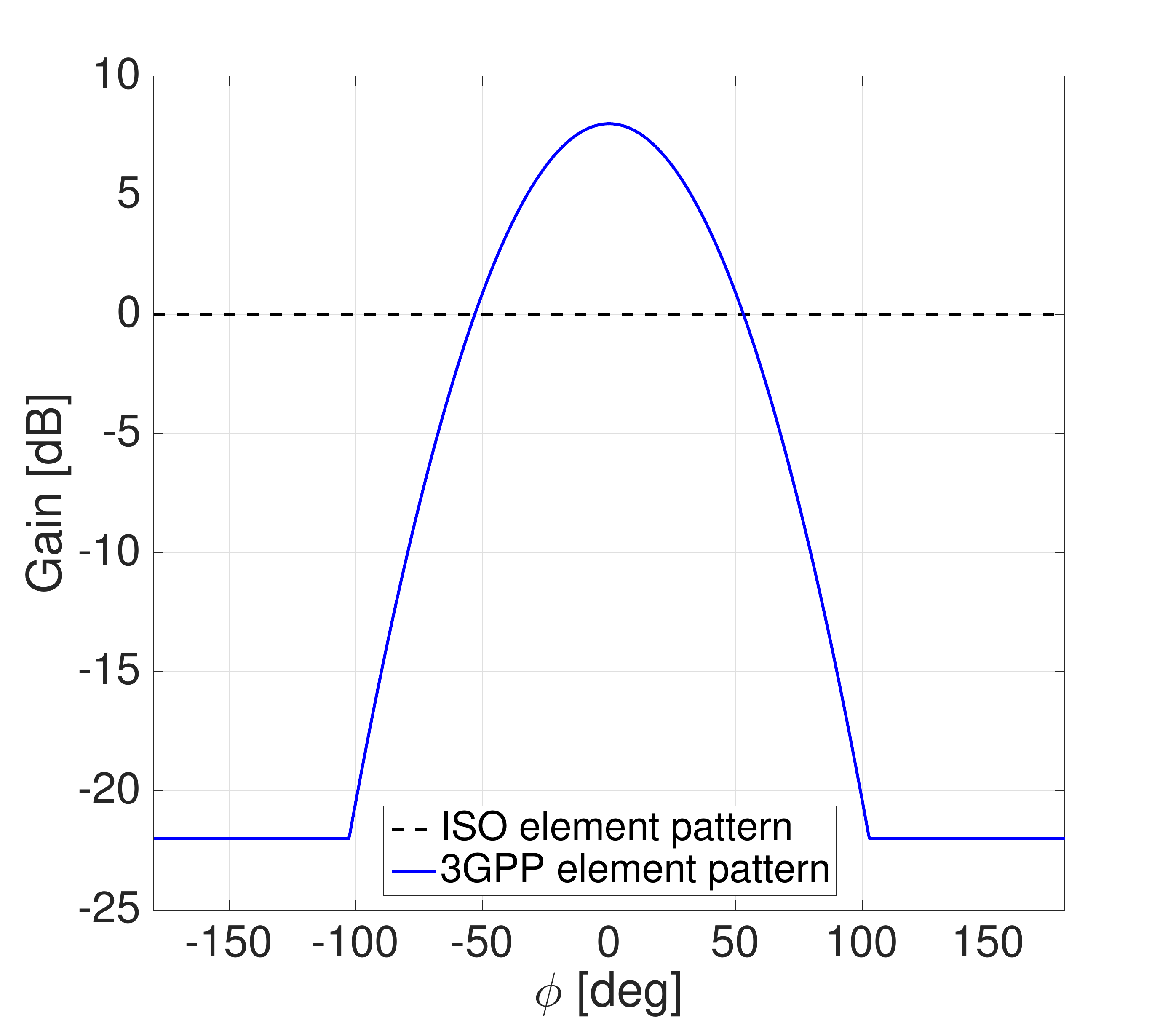}
\caption{Element radiation gain.}
\label{example_element_radiation_pattern}
\end{subfigure}
\hfill
\begin{subfigure}[b]{0.45\textwidth}
\centering
\includegraphics[width=1\textwidth]{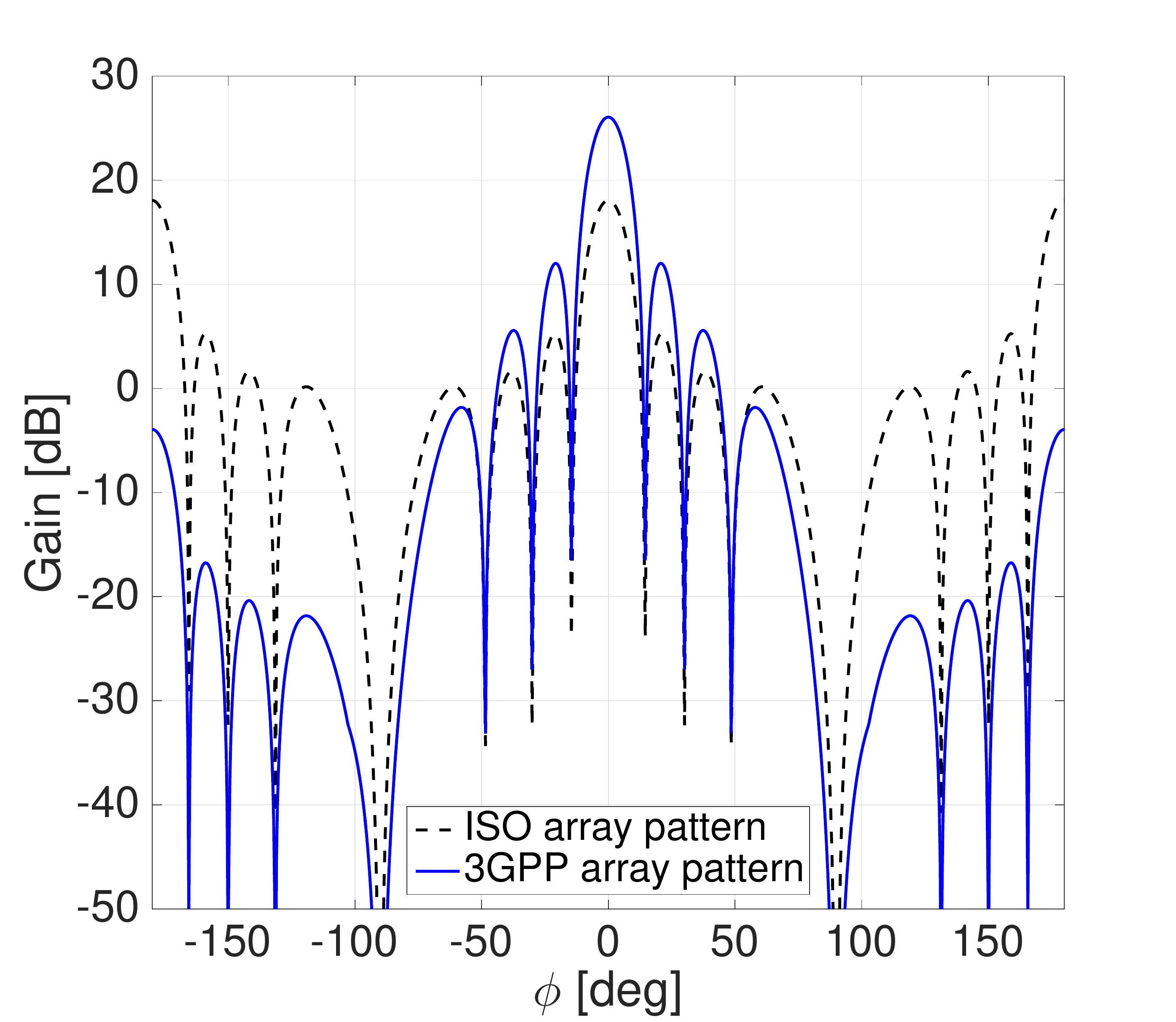}
\caption{Array radiation gain with $64$ antenna elements.}
\label{example_array_radiation_pattern}
\end{subfigure}
\caption{Illustrations of the element radiation gains and the array radiation gains for the $\ISO$ and $\3G$ element patterns, with respect to the horizontal steering angle $\phi\in[-180^{\circ},180^{\circ}]$ while the vertical steering angle $\theta$ is fixed at $90^\circ$.}
\label{global_patterns_view}
\end{figure*} 
The $\3G$ element pattern is realized according to the specifications in~\cite{mmWave_3gpp_channel,antenna_3gpp} and~\cite{3d_channel_3gpp}.
First, differently from the previous configuration, it implies the use of three sectors, thus three arrays, placed as in traditional mobile networks\footnote{We note that, even if three sectors are present in each BS site, only a single sector is active and transmitting in each time instant.}.
Second, the single element radiation pattern presents high directivity with a maximum gain in the main-lobe direction of about 8~dBi. 
The $\3G$ $A_E$ of each single antenna element is composed of horizontal and vertical radiation patterns.
Specifically, this last pattern $A_{E,V}(\theta)$ is obtained as
\begin{align}
A_{E,V}(\theta) = - \min \left\{ 12 \left( \frac{\theta - 90}{\theta_{3 \text{dB}}}\right)^2, SLA_V \right\},
\end{align}
where $\theta_{3 \text{dB}} = 65^\circ$ is the vertical 3~dB beamwidth, and $ SLA_V = 30 \text{ dB}$ is the side-lobe level limit.
Similarly, the horizontal pattern is computed as 
\begin{align}
A_{E,H}(\phi) = - \min \left\{ 12 \left( \frac{\phi}{\phi_{3 \text{dB}}}\right)^2, A_m \right\},
\end{align}
where $\phi_{3 \text{dB}} = 65^\circ$ is the horizontal 3~dB beamwidth, and $ A_m = 30$~dB is the front-back ratio.
Using together the previously computed vertical and horizontal patterns we can compute the 3D antenna element gain for each pair of angles as
\begin{align}
A_E^{(\3G)}(\theta,\phi) = G_{\max}- \min \left\{- \left[ A_{E,V}(\theta) + A_{E,H}(\phi) \right], A_m \right\},
\label{global_pattern_equation}
\end{align}
where $G_{\max} = 8$~dBi is the maximum directional gain of the antenna element~\cite{mmWave_3gpp_channel}.
The expression in~\eqref{global_pattern_equation} provides the dB gain experienced by a ray with angle pair $(\theta,\phi)$ due to the effect of the element radiation pattern.

\subsubsection{Array radiation pattern} 
The antenna array radiation pattern $A_A^{(z)}(\theta,\phi)$ determines how much power is radiated from an antenna array towards the steering direction $(\theta,\phi)$. Following~\cite{antenna_3gpp}, the array radiation pattern with a given element radiation pattern $A_E^{(z)}(\theta,\phi)$ is provided as
\begin{align}
A_A^{(z)}(\theta,\phi) = A_E^{(z)}(\theta,\phi) + \mathsf{AF}(\theta,\phi).
\label{ralation_array_element}
\end{align}
The last term $\mathsf{AF}(\theta,\phi)$ is the array factor with the number $n$ of antenna elements, given as
\begin{align}
\mathsf{AF}(\theta,\phi) = 10 \log_{10}\left[ 1+ \rho \left( \left| \textbf{a} \cdot \textbf{w}^T \right|^2-1\right) \right],
\label{array_factor}
\end{align}
where $\rho$ is the correlation coefficient, set to unity by assuming the same correlation level between signals in all the transceiver paths~\cite{antenna_3gpp}.
Since it represents the physical specifications of the array, $\mathsf{AF}$ is equally computed for both $\ISO$ and $\3G$ models.
The term $\textbf{a} \in \mathbb{C}^{n}$ is the amplitude vector, set as a constant $1/\sqrt{n}$ while assuming that all the antenna elements have equal amplitude.
The term $\textbf{w} \in \mathbb{C}^{n}$ is the beamforming vector, which includes the mainlobe steering direction, to be specified in Section~\ref{aligned_and_misaligned_gains}.
This last term depends on the considered pair of angles $(\theta,\phi)$, although, for ease of notation, we are not reporting this dependency in the equation. 
Further explanation of the relation between array and element patterns can be found in~\cite{rebato18} and~\cite{antenna_3gpp}.

In Fig.~\ref{example_element_radiation_pattern} we report a comparison of the two continuous element radiation patterns (i.e., $A_E$).
The figure permits to understand the difference between the $\ISO$ element pattern showing a fixed gain and the $\3G$ element pattern providing $8$ dBi directivity.
As a consequence of the element pattern used, we can see the respective shape of the array radiation pattern (i.e., $A_A$) in Fig.~\ref{example_array_radiation_pattern}. 
The plot permits to see the reduction of undesired sidelobes and backward propagation when considering the $\3G$ curve with respect to the $\ISO$ element pattern.
Furthermore, shape and position of the main and undesired lobes vary as a function of the steerable direction.
Further definitions and accurate examples for these concepts can be found in~\cite{rebato18}.

\subsubsection{Field pattern (i.e., antenna gain)}
Finally, applying the given antenna array pattern $A_A^{(z)}(\theta,\phi)$, we obtain the antenna gain for the channel computations.
This gain consists of a vertical field pattern $\mathsf{F}^{(z)} (\theta,\phi)$ and a horizontal field pattern $\mathsf{G}^{(z)} (\theta,\phi)$, with the polarization slant angle $\zeta$.
For simplicity, in this study we consider a purely vertically polarized antenna, i.e., $\zeta = 0$.
Following~\cite{3d_channel_3gpp}, the vertical and horizontal field patterns are thereby given as follows
\begin{align}
\label{field_factor_expression}
\mathsf{F}^{(z)} (\theta,\phi) &= \sqrt{A_A^{(z)}(\theta,\phi)}\cos (\zeta)= \sqrt{A_A^{(z)}(\theta,\phi)},\\
\mathsf{G}^{(z)} (\theta,\phi) &= \sqrt{A_A^{(z)}(\theta,\phi)}\sin (\zeta) = 0.
\end{align} 

\subsection{Channel gain}
\label{propagation_and_channel_gain}
Following the system-level simulator settings \cite{akdeniz14}, we divide the channel gains into two parts: (i) \emph{path loss} that depends on the link distance; and (ii) the channel gain multiplicative component.
The latter gain is affected not only by the channel randomness but also by the antenna array directions.
The following channel gain computation aspects are independent of the different radiation pattern considered, thus they are valid for both $\ISO$ and $\3G$.

The antenna array direction is determined by the BS-UE association.
To elaborate, for each associated BS-UE link, denoted as the desired link, their beam directions are aligned, pointing their main-lobe centers towards each other.
As a consequence, for all non-associated BS-UE links, denoted as interfering links, the beam directions can be misaligned.
In order to distinguish them in (ii), we define \emph{aligned gain} and \emph{misaligned gain} as the channel gain for the desired link and for an interfering link, respectively. 
The definitions of path loss and aligned/misaligned gains are specified in the following subsections.

\subsubsection{Path loss}
By definition, the set of BS locations follows a homogeneous Poisson point process (HPPP) $\Phi$ with density $\lambda_b$.
At the typical UE, the desired/interfering links can be in either LoS or NLoS state.
To be precise, from the perspective of the typical UE, the set $\Phi$ of all the BSs is partitioned into a set of LoS BSs $\Phi_L$ and a set of NLoS BSs~$\Phi_N$. According to the minimum path loss association rule, the desired link can be either LoS or NLoS, specified by using the subscript $i\in\{L,N\}$.
Likewise, the LoS/NLoS state of each interfering link is identified by using the subscript $j\in\{L,N\}$.

For a given link distance~$r$, the LoS and NLoS state probabilities are $p_L(r) =e^{-0.0149r}$ and $p_N(r) = 1-p_L(r)$~\cite{akdeniz14,samimi15,bai15}.
Here, compared to the system-level simulator settings in~\cite{akdeniz14,samimi15}, we neglect the outage link state induced by severe distance attenuation.
This assumption does not incur a loss of generality for our $\SINR$ analysis, since the received signal powers that correspond to outage are typically negligibly small.

When a connection link has distance $r$ and is in state $j\in\{L,N\}$, transmitted signals passing through this link experience the following path loss attenuation
\begin{align}
\ell^j(r) = \beta_j r^{-\alpha_j},
\label{Eq:pathloss}
\end{align}
where $\alpha_j$ indicates the path loss exponent and $\beta_j$ is the path loss gain at unit distance~\cite{akdeniz14,Rappaport:17}. 

\subsubsection{Aligned and misaligned gains}
\label{aligned_and_misaligned_gains}
In both $\ISO$ and $\3G$ element patterns, for a given link, a random channel gain is determined by the NYU channel model that follows mmWave channel specific parameters~\cite{akdeniz14,samimi15} based on the WINNER~II model~\cite{winner2}.
These parameters are summarized in Tab.~\ref{Table:Notations}, and discussed in the following subsections.
In this model, each link comprises $K$ clusters that correspond to macro-level scattering paths.
For cluster $k\leq K$, there exist $L_k$ subpaths, as visualized in Fig.~\ref{globalview}.
Moreover, the first cluster angle (i.e., $\phi_k, k=1$) exactly matches the LOS direction between transmitter and receiver in the simulated link.
\begin{table}
\centering
\caption{List of notations and channel parameters considered in the NYU mmWave network simulator~\cite{mezzavilla15}.}
\small
\renewcommand{\arraystretch}{1.5}
\resizebox{\columnwidth}{!}{\begin{tabular}{r  l }
\toprule
\hspace{-10pt}\bf{Notation} &\hspace{-5pt} \textbf{Meaning}: Parameters\\
\cmidrule(r){1-1}\cmidrule(r){2-2}
\hspace{-10pt} $f$ & Carrier frequency: 28 GHz\\
\hspace{-10pt} $\Phi_b$ & BS locations following a HPPP with density $\lambda_b$\\
\hspace{-10pt} $p_L(r)$ & LoS state probability at distance $r$: $p_L=e^{-0.0149r}$\\
\hspace{-10pt} $x_o,\; x$ & Serving and interfering BSs or their coordinates\\
\hspace{-10pt} $\alpha_j$ & Path loss exponent, with $j\in\{L,N\}$: $\alpha_L=2$, $\alpha_N=2.92$\\
\hspace{-10pt} $\beta_j$ & Path loss gain at unit distance: $\beta_L=10^{-7.2}$, $\beta_N=10^{-6.14}$\\
\hspace{-10pt} $\ell^j(r)$ & Path loss at distance $r$ in LoS/NLoS state\\
\hspace{-10pt} $n_\text{TX},\; n_\text{RX}$ & \# of antennas of a BS and a UE\\
\hspace{-10pt} \scriptsize$G_o^{(z)},\; G_x^{(z)}$\normalsize & Aligned and misaligned gains, with $z\in\{\ISO,\3G\}$ \\
\hspace{-10pt} $f_{G_o}^{(z)}, f_{G_x}^{(z)}$ & Aligned and misaligned gain PDFs\\
\hspace{-10pt} $K$ & \# of clusters $\sim \max\{\textsf{Poiss}(1.8),1\}$ \\
\hspace{-10pt} $L_k$ & \# of subpaths in the $k$-th cluster $\sim \textsf{DiscreteUni}[1,10]$ \\
\hspace{-10pt} $\phi_{kl}^{\text{RX}}$, $\phi_{kl}^{\text{TX}}$ & Angular spread of subpath $l$ in cluster $k$~\cite{akdeniz14}: \\[-.2em]
\hspace{-10pt} & $\phi_{k}^{(\cdot)}\hspace{-3pt}\sim \textsf{Uni}[0,2\pi], \forall k \ne 1,$ $\phi_{kl}^{(\cdot)}=\phi_{k}^{(\cdot)}\hspace{-3pt}+ (-1)^l s_{kl}/2$\\[-.2em]
\hspace{-10pt} & $s_{kl}\sim \max\{\textsf{Exp}(0.178),0.0122\},$\\
 \hspace{-10pt} $g_{kl}$ & Small-scale fading gain: $g_{kl}=\sqrt{P_{kl}}\exp(-j2\pi \tau _{kl}f)$\\
 \hspace{-10pt} $\tau _{kl}$& Delay spread induced by different subpath distances.\\
\hspace{-10pt} $P_{kl}$ & Power gain of subpath $l$ in cluster $k$~\cite{samimi15}:\\
\hspace{-10pt} & $U_k \sim \textsf{Uni}[0,1]$, $Z_k \sim \mathcal{N}(0,4^2)$, $V_{kl} \sim \textsf{Uni}[0,0.6]$,\\[-.2em]
\hspace{-10pt} & $P_{kl}={P_{kl}^\prime}/{\sum P_{kl}^\prime}$, $P_{kl}^\prime = {U_k^{\tau_{kl}-1}10^{-0.1 Z_k+V_{kl}}}/{L_k}$, $\tau_{kl}=2.8$
\\
\bottomrule
\end{tabular}}
\label{Table:Notations}
\end{table}

Given a set of clusters and subpaths, the channel matrix of each link is represented as
\begin{align}
\textbf{H}^{(z)}= \sum_{k=1}^{K}\sum_{l=1}^{L_k}g_{kl} \mathsf{F}^{(z)}_{\text{RX}}\left(\phi^{\text{RX}}_{kl}\right) \textbf{u}_{\text{RX}}\left(\phi^{\text{RX}}_{kl}\right) \mathsf{F}^{(z)}_{\text{TX}}\left(\phi^{\text{TX}}_{kl}\right) \textbf{u}^*_{\text{TX}}\left(\phi^{\text{TX}}_{kl}\right)
\label{channel_matrix}
\end{align}
where {$g_{kl}$} is the small-scale fading gain of subpath $l$ in cluster $k$, and $\textbf{u}_{\text{RX}}$ and $\textbf{u}_{\text{TX}}$ are the 3D spatial signature vectors of the receiver and transmitter, respectively. Note that $\textbf{u}^*_{\text{TX}}$ stands for the complex conjugate of vector $\textbf{u}_{\text{TX}}$.
Furthermore, for brevity, we use subscript or superscript TX (RX), referring to a transmitter (receiver) related term.
Moreover, $\phi^{\text{RX}}_{kl}$ is the angular spread of horizontal angles of arrival (AoA) and  $\phi^{\text{TX}}_{kl}$ is the angular spread of horizontal angles of departure (AoD), both for subpath $l$ in cluster $k$~\cite{akdeniz14}.
Note that, for ease of computation, we consider a planar network and channel, i.e., we neglect vertical signatures by setting their angles to 90$^\circ$ (i.e., ${\pi}/{2}$ radian).
Finally, $\mathsf{F}_{\text{TX}}^{(z)}$ and $\mathsf{F}_{\text{RX}}^{(z)}$ are the field factor terms of transmitter and receiver antennas, respectively and they are computed as in~\eqref{field_factor_expression} with $z\in\{\ISO,\3G\}$. 

We consider directional beamforming where the mainlobe center of a BS's transmit beam points at its associated UE (we recall that $\phi_1$ is the mainlobe center angle as shown in the channel illustration of Fig.~\ref{globalview}), while the mainlobe center of a UE's receive beam aims at the serving BS.
We assume that both beams can be steered in any directions.
Therefore, considering the $\ISO$ element pattern, we can generate a beamforming vector for any possible angle in $\left[0,360^{\circ}\right]$.
Instead, with the three-sectors consideration adopted in the $\3G$ element pattern, the beamforming vectors for any possible angles are mapped within one of the three sectors, thus using an angle in the interval $\left[0,120^{\circ}\right]$.

At the typical UE, the aligned gain $G_o^{(z)}$ is its beamforming gain towards the serving BS at $x_o$.
With a slight abuse of notation for the subscript $x_o$, $G_o^{(z)}$ is represented as
\begin{align}
& G_o^{(z)} = |\textbf{w}^T_{\text{RX}_{x_o}} \textbf{H}^{(z)}_{x_o} \textbf{w}_{\text{TX}_{x_o}} |^2\\
&\hspace{-7pt}= \left|  \sum_{k=1}^{K}\sum_{l=1}^{L_k}g_{kl} \mathsf{F}^{(z)}_{\text{RX}} \left( \textbf{w}^T_{\text{RX}_{x_o}}\textbf{u}_{\text{RX}_{x_o}} \right) \mathsf{F}^{(z)}_{\text{TX}} \left( \textbf{u}^*_{\text{TX}_{x_o}}\textbf{w}_{\text{TX}_{x_o}} \right) \right|^2
\label{bf_gain_1}
\end{align}
where $\textbf{w}_{\text{TX}_{x_o}} \in \mathbb{C}^{n_{\text{TX}}}$ is the transmitter beamforming vector and $\textbf{w}^T_{\text{RX}_{x_o}} \in \mathbb{C}^{n_{\text{RX}}}$ is the transposed receiver beamforming vector computed as in~\cite{tse_book,rebato18}.
Their values contain information about the mainlobe steering direction and both are computed using the first cluster angle $\phi_1$ as
\begin{align}
\textbf{w}^T_{\text{TX}} &= [w_{1,1},w_{1,2},  \dots,w_{\sqrt{n_{\text{TX}}},\sqrt{n_{\text{TX}}}}],
\end{align}
where $w_{p,r} = \exp\( j 2 \pi \[ (p-1) {\Delta_V} \Psi_p /{\lambda} + (r-1) {\Delta_H} \Psi_r/{\lambda} \] \)$, for all $p,r \in \{1,\dots,\sqrt{n_{\text{TX}}}\}$, $\Psi_p = \cos \left(\theta_s\right)$, and $\Psi_r = \sin \left(\theta_s\right) \sin \left(\phi_1\right)$.
The terms $\Delta_V$ and $\Delta_H$ are the spacing distances between the vertical and horizontal elements of the array, respectively.
Then, angles $\theta_s$ and $\phi_s$ are the steering angles and $\theta_s$ is kept fixed to $90^\circ$.
We assume all elements to be evenly spaced on a two-dimensional plane, thus it equals $\Delta_V = \Delta_H = {\lambda}/{2}$. The same expression can be used to compute the receiver beamforming vector $\textbf{w}_{\text{RX}}$ with the exception that its dimension is $n_{\text{RX}}$.

Similarly, the typical UE's misaligned gain $G_x^{(z)}$ is its beamforming gain with an interfering BS at $x$
\begin{align}
&G_{x}^{(z)} = |\textbf{w}^T_{\text{RX}_{x}} \textbf{H}^{(z)}_{x} \textbf{w}_{\text{TX}_{x}} |^2
\label{bf_gain_2}
\end{align}
where $\textbf{w}_{\text{TX}_{x}}$ and $\textbf{w}_{\text{RX}_{x}}$ respectively are the transmitter and receiver beamforming vectors.
It is noted that both $G_o^{(z)}$ and $G_x^{(z)}$ incorporate the effects not only of the mainlobes but also of all the other sidelobes.
We highlight that even if both aligned and misaligned gain definitions are valid for both the $\ISO$ and $\3G$ configurations, the gains will have a different distribution in the two radiation patterns. 

\subsection{$\SINR$ definition}
\label{sinr_definition}
The typical UE is regarded as being located at the origin, which does not affect its $\SINR$ behaviors thanks to Slivnyak's theorem~\cite{HaenggiSG} under the HPPP modeling of the BS locations.
At the typical UE, let $x_o$ and all the $x\in \Phi_i$ respectively indicate the associated and interfering BSs as well as their coordinates.
We note that the set $\Phi_i$ represents BS locations following a HPPP with density $\lambda_i, i\in \{L,N\}$.

Using equations~\eqref{Eq:pathloss},~\eqref{bf_gain_1}, and~\eqref{bf_gain_2}, we can represent $\SINR_i$ as the received $\SINR$ at the typical UE associated with $x_o\in\Phi_i, i\in \{L,N\}$, which is given by
\begin{align}
\SINR_i &= \frac{ G_o^{(z)} \ell_i(r_{x_o}^i) }{\sum\limits_{x\in \Phi_L /x_o } G^{(z)}_{x} \ell_L(r_x^L) + \sum\limits_{x\in \Phi_N /x_o } G_{x}^{(z)} \ell_N(r_x^N) + \sigma^2},
\label{equation_sinr}
\end{align}
where the term $r_{x_o}^i$ denotes the association distance of the typical UE associating with $x_o\in\Phi_i$ and along similar lines, $r_{x}^i$ denotes the association distance of a generic UE associating with $x\in\Phi_i$ and $i \in \{L,N\}$.
Knowing that the typical UE is located in the origin $o$, $r_x$ is equals to $\|x\|$.
Here, we assume that each BS transmits signals with the maximum power $P_\tx$ through the bandwidth $W$. In \eqref{equation_sinr}, $\SINR_i$ is normalized by $P_\tx$. The term $\sigma^2$ denotes the normalized noise power $\sigma^2$ that equals $\sigma^2 = W N_{0}/P_{\text{TX}}$ where $N_0$ is the noise spectral density per unit bandwidth.

\section{Aligned and Misaligned Gain Distributions}
\label{Gain_distribution}
Starting from the expressions derives in the previous section, it is practically infeasible to further approximate aligned and misaligned gains using analytic methods, as analyzing each of their subordinate terms is a major task in itself, as shown by related works.
Therefore, in this section we focus on the aligned gain $G_o^{(z)}$ in~\eqref{bf_gain_1} and the misaligned gain $G_{x}^{(z)}$ in~\eqref{bf_gain_2} with $\ISO$ and $\3G$ element patterns, and aim at deriving their distributions.

Following the definitions in Sect.~\ref{propagation_and_channel_gain}, the aligned gain $G_o^{(z)}$ is obtained for the desired received signal when the angles of the beamforming vectors $\textbf{w}_{\text{TX}_{x_o}}$ and $\textbf{w}_{\text{RX}_{x_o}}$ are aligned with the AoA and AoD of the spatial signatures $\textbf{u}_{\text{TX}_{x_o}}$ and $\textbf{u}_{\text{RX}_{x_o}}$ in the channel matrix $\textbf{H}^{(z)}_{x_o}$. The misaligned gain $G_x^{(z)}$ is calculated for each interfering link with the beamforming vectors and spatial signatures that are not aligned.\footnote{At the typical UE, the serving BS's beamforming is aligned with the typical UE, whereas the beamforming vectors of interfering BSs are determined by their own associated UEs that are uniformly distributed. For this reason, each interfering BS's beamforming has a circularly uniform orientation. Consequently, in~\eqref{bf_gain_2}, the angles of the beamforming vectors $\textbf{w}_{\text{TX}_x}$ and $\textbf{w}_{\text{RX}_x}$ as well as the angles of the spatial signatures $\textbf{u}_{\text{TX}}$ and $\textbf{u}_{\text{RX}}$ are not aligned with the angles of $G_o^{(z)}$, which are independent and identically distributed (i.i.d.) across different interfering BSs.}
Fig.~\ref{globalview} shows an example of misalignment between the beam of the desired signal (yellow or green colored beam) and the interfering BSs beams (red colored beams). 

In the following subsections, using curve fitting with the system-level simulation, we derive the distributions of the aligned gain $G_o^{(z)}$ and the misaligned gain $G_x^{(z)}$.

\subsection{Aligned gain distribution}
\label{alignedt_gain_distribution}
\begin{figure}[t!]
\centering
\setlength{\belowcaptionskip}{-0.4cm}
\includegraphics[width=\columnwidth]{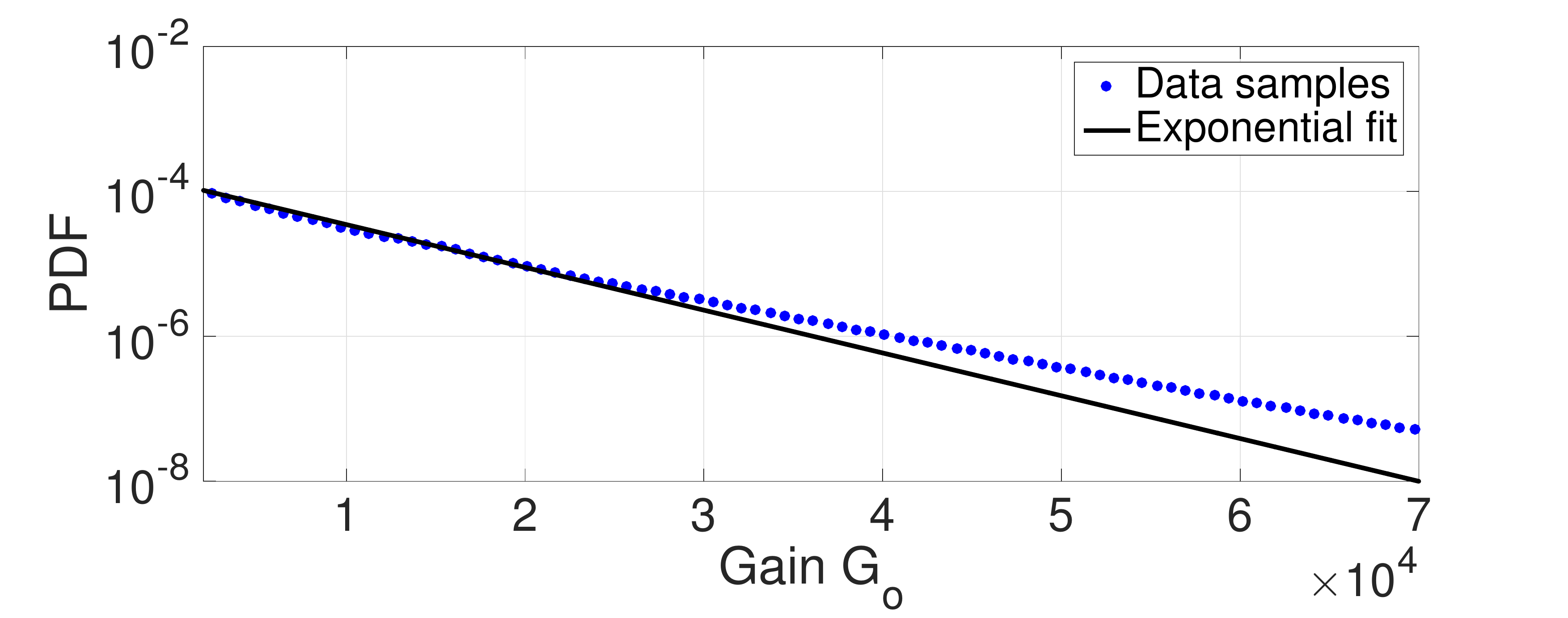}
\caption{Fitting of the aligned gain $G_o^{(\ISO)}$ with the $\ISO$ element pattern. The empirical PDF of $G_o^{(\ISO)}$ is obtained by the NYU mmWave network simulator~\cite{mezzavilla15}, and is fit with the \emph{exponential} distribution in Remark~1 ($n_{\text{RX}} = 64$, $n_{\text{TX}} = 256$).}
\label{exponential_fit_G0}
\end{figure}

Running a large number of independent runs of the NYU simulator we empirically evaluated the distribution of the aligned gain $G_o^{(z)}$.
From the obtained data samples we have noticed that $G_o^{(z)}$ is roughly exponentially distributed $G_o^{(\ISO)} \sim \mathsf{Exp}(\mu_o)$ when an $\ISO$ element pattern is used.
Indeed, the signal's real and imaginary parts are approximately independent and identically distributed zero-mean Gaussian random variables.
This exponential behavior finds an explanation in the small-scale fading effect implemented in the channel model using the power gain term $P_{kl}$ computed as reported in Tab.~\ref{Table:Notations}.
We report in Fig.~\ref{exponential_fit_G0} an example of the exponential fit of the simulated distribution.
The fit has been obtained using the \emph{curve fitting toolbox} of MATLAB.

\begin{figure}
\setlength{\belowcaptionskip}{-0.5cm}
\centering
\includegraphics[width=\columnwidth]{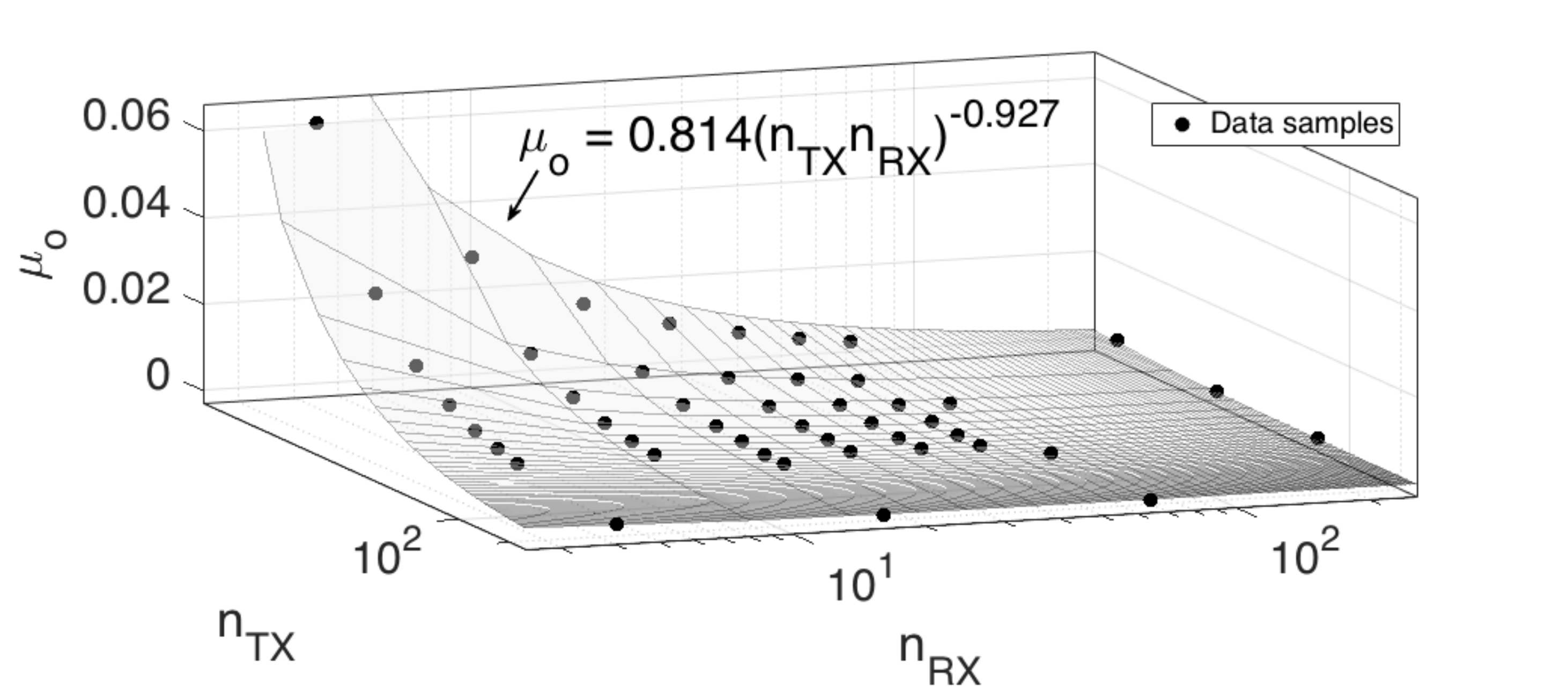}
\caption{Fitting of the aligned gain distribution parameter $\mu_o$ with the $\ISO$ element pattern, with respect to the number of antenna elements $n_{\text{TX}}$ and~$n_{\text{RX}}$.}
\label{power_fit}
\end{figure}

For the purpose of deriving an analytical expression, it is also interesting to evaluate the behavior of $\mu_o$ as a function of the number of antenna elements at both receiver and transmitter sides. 
For this reason, in our analysis we consider the term $\mu_o$ as a function of the number of elements.
We show in Fig.~\ref{power_fit} the trend of the parameter $\mu_o$ versus the number of antenna elements at the transmitter side $n_{\text{TX}}$ and at the receiver side $n_{\text{RX}}$.
Again, using the MATLAB curve fitting toolbox, we have obtained a two-dimensional power fit where the value of $\mu_o$ can be obtained as in the following remark.

\begin{remark} (Aligned Gain, $\ISO$) \emph{At the typical UE, under the $\ISO$ antenna model, the aligned gain $G_o^{(\ISO)}$  can be approximated by an \emph{exponential} distribution with probability density function (PDF)
\begin{align}
f_{G_o}^{(\ISO)}(y;\mu_o) = \mu_o e^{-\mu_o y} \label{Eq:GoISO}
\end{align}
where $\mu_o = \frac{0.814} { (n_{\emph{\text{TX}}}n_{\emph{\text{RX}}})^{0.927}}$.}
\end{remark}
This result provides a fast tool for future calculations.
Indeed, the expression found for the gain permits to avoid running a detailed simulation every time.
We note that from a mathematical point of view the surface of the term $\mu_o (n_{\text{TX}},n_{\text{RX}})$ is symmetric.
In fact, the gain does not depend individually on the number of antennas at the transmitter or receiver sides, but rather on their product, so we can trade the complexity at the BS for that at the UE if needed.

\begin{figure}[t!]
\centering
\setlength{\belowcaptionskip}{-0.4cm}
\includegraphics[width=\columnwidth]{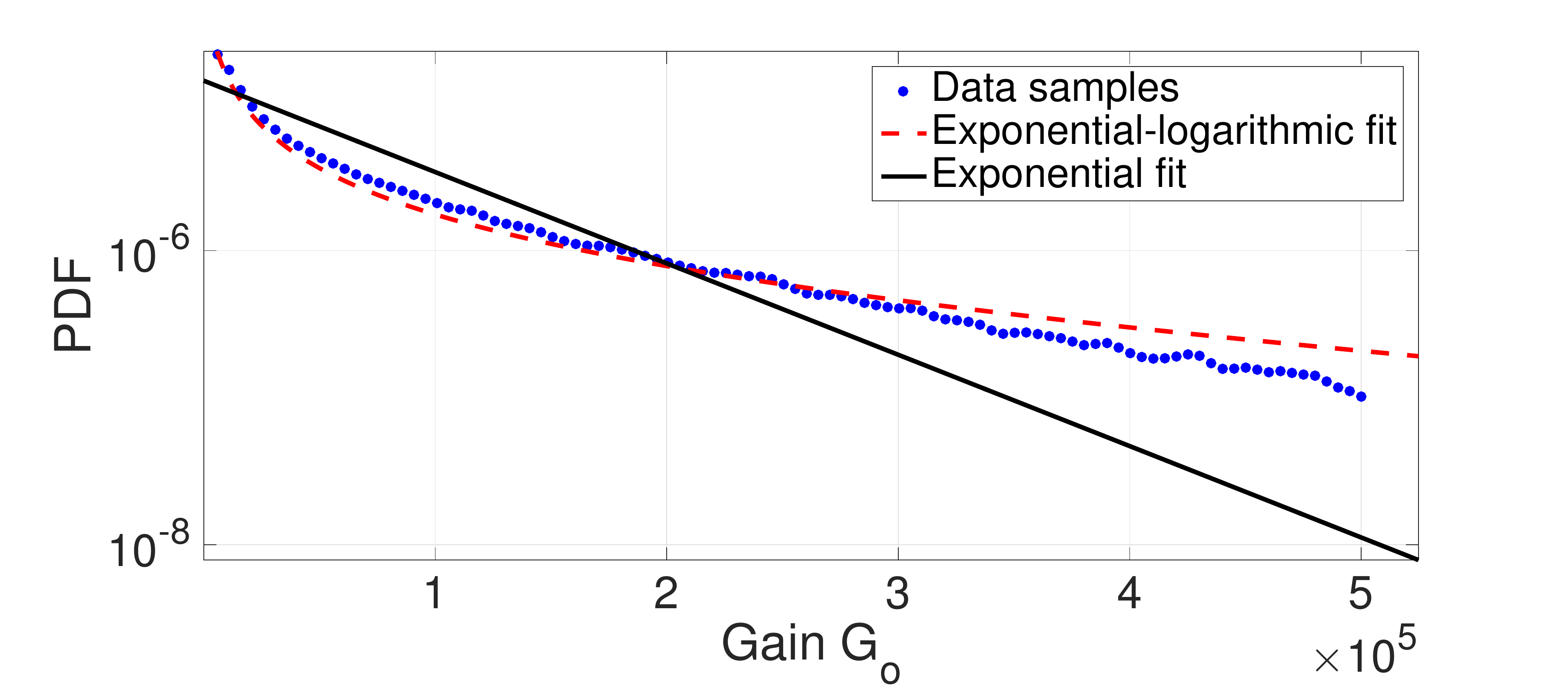}
\caption{Fitting of the aligned gain $G_o^{(\3G)}$ with the $\3G$ element pattern. The empirical PDF of $G_o^{(\3G)}$ fits with the \emph{exponential-logarithmic} distribution in Remark~2. It no longer fits with an exponential distribution, as opposed to the $\ISO$ element pattern's ($n_{\text{RX}} = 64$, $n_{\text{TX}} = 256$).}
\label{comparison_dist_g0}
\end{figure}

By contrast, using the $\3G$ element pattern, we have noticed that the data samples of $G_o^{(\3G)}$ can no longer be approximated as an exponentially distributed random variable. Instead, an \emph{exponential-logarithmic} distribution provides the most accurate fitting result with the simulated desired gain, validated by simulation as shown in Fig.~\ref{comparison_dist_g0}. 

\begin{remark} (Aligned Gain, $\3G$) \emph{At the typical UE, and adopting the $\3G$ element pattern, the aligned gain $G_o^{(\3G)}$ can be approximated by an \emph{exponential-logarithmic} distribution with PDF
\begin{align}
f_{G_o}^{(\3G)}(y; b_o, p_o) = \frac{1}{- \ln (p_o)} \frac{b_o(1-p_oe^{-b_o y})}{1-(1-p_o)e^{-b_o y}}.
\end{align}
where the parameters $b_o$ and $p_o$ are specified in Tab.~\ref{table_beta_p}.}
\end{remark}

Exponential-logarithmic distributions are often used in the field of reliability engineering, particularly for describing the lifetime of a device with a decreasing failure rate over time~\cite{Tahmasbi:08}.
Its tail is lighter than that of the exponential distribution, which is explained by the $\3G$ element pattern's high directivity and sidelobe attenuation that mostly yield a higher aligned gain than the $\ISO$ element pattern's aligned gain.

An exponential-logarithmic distribution is determined by using two parameters $b_o$ and $p_o$, as opposed to the $\ISO$ element pattern's exponential distribution with a single parameter $\mu_o$.
Precisely, the distribution is given by a random variable that is the minimum of $N$ independent realizations from $\text{Exp}(b_o)$, while $N$ is a realization from a logarithmic distribution with parameter $1-p_o$.
Due to its generation procedure, the relationship between the two parameters and the number of antenna elements is not representable with a simple function in a way to be generalized as done in Remark~1 for the $\ISO$ element pattern.
In particular, due to the extreme characteristics of the gain, even a small variation in the well-fitted parameters yields a significant change in the fitting accuracy.
For this reason, obtaining a good-fit of the parameters that can be generalized requires an exhaustive search, with an extremely large number of combinations.
Therefore, for $16$ practically possible combinations of $n_\text{TX}$ and $n_{\text{RX}}$, the appropriate values of $b_o$ and $p_o$ are provided in Tab.~\ref{table_beta_p} by curve-fitting of the system-level simulation results.

\begin{table}
\caption{Aligned gain's \emph{exponential-logarithmic} distribution parameters $(b_o,p_o)$ with the $\3G$ element pattern for different $n_{\text{TX}}$ and $n_{\text{RX}}$. The table is symmetric, so we hereafter report only the upper triangular part.}
\centering
\renewcommand{\arraystretch}{1.2}
\resizebox{\columnwidth}{!}{\begin{tabular}{ cc||cccc}
\toprule
\multicolumn{2}{c||}{\multirow{2}{*}{$(b_o,p_o)$}}                                           & \multicolumn{4}{c}{$n_{\text{TX}}$}                                                                                                                                                                           \\ \cline{3-6} 
\multicolumn{2}{c||}{}      & \multicolumn{1}{c|}{\textbf{4}}                                                                           & \multicolumn{1}{c|}{\textbf{16}}                                                                          & \multicolumn{1}{c|}{\textbf{64}}                                                                          & \textbf{256}                                                                         \\ \hline \hline
\multicolumn{1}{c|}{\multirow{4}{*}{$n_{\text{RX}}$}} & \textbf{4}   & \multicolumn{1}{c|}{\begin{tabular}[c]{@{}l@{}}$(0.002,$ $0.112)$\end{tabular}} & \multicolumn{1}{c|}{\begin{tabular}[c]{@{}l@{}}$(4\text{e-}4,$ $0.075)$\end{tabular}} & \multicolumn{1}{c|}{\begin{tabular}[c]{@{}l@{}}$(0.0001,$ $0.0713)$\end{tabular}} & \begin{tabular}[c]{@{}l@{}}$(7.84\text{e-}5,$ $0.15)$\end{tabular} \\ \cline{2-6} 
\multicolumn{1}{c|}{}                                 & \textbf{16}  & \multicolumn{1}{c|}{\begin{tabular}[c]{@{}l@{}}$-$\end{tabular}} & \multicolumn{1}{c|}{\begin{tabular}[c]{@{}l@{}}$(2\text{e-}4,$ $0.15)$\end{tabular}} & \multicolumn{1}{c|}{\begin{tabular}[c]{@{}l@{}}$(8.24\text{e-}5,$ $0.511)$\end{tabular}} & \begin{tabular}[c]{@{}l@{}}$(1.93\text{e-}5,$ $0.1223)$\end{tabular} \\ \cline{2-6} 
\multicolumn{1}{c|}{}                                 & \textbf{64}  & \multicolumn{1}{c|}{\begin{tabular}[c]{@{}l@{}}$-$\end{tabular}} & \multicolumn{1}{c|}{\begin{tabular}[c]{@{}l@{}}$-$\end{tabular}} & \multicolumn{1}{c|}{\begin{tabular}[c]{@{}l@{}}$(1.84\text{e-}5,$ $0.15)$\end{tabular}}  & \begin{tabular}[c]{@{}l@{}}$(4.83\text{e-}6,$ $0.089)$\end{tabular} \\ \cline{2-6} 
\multicolumn{1}{c|}{}                                 & \textbf{256} & \multicolumn{1}{c|}{\begin{tabular}[c]{@{}l@{}}$-$\end{tabular}} & \multicolumn{1}{c|}{\begin{tabular}[c]{@{}l@{}}$-$\end{tabular}} & \multicolumn{1}{c|}{\begin{tabular}[c]{@{}l@{}}$-$\end{tabular}} & \begin{tabular}[c]{@{}l@{}}$(1.96\text{e-}6,$ $0.1126)$
\end{tabular}
\\
\bottomrule
\end{tabular}}
\label{table_beta_p}
\end{table}

\subsection{Misaligned gain distribution}
\label{beam_misalignment_gain_distribution}
\begin{figure}[t!]
\centering
\setlength{\belowcaptionskip}{-0.4cm}
\includegraphics[width=\columnwidth]{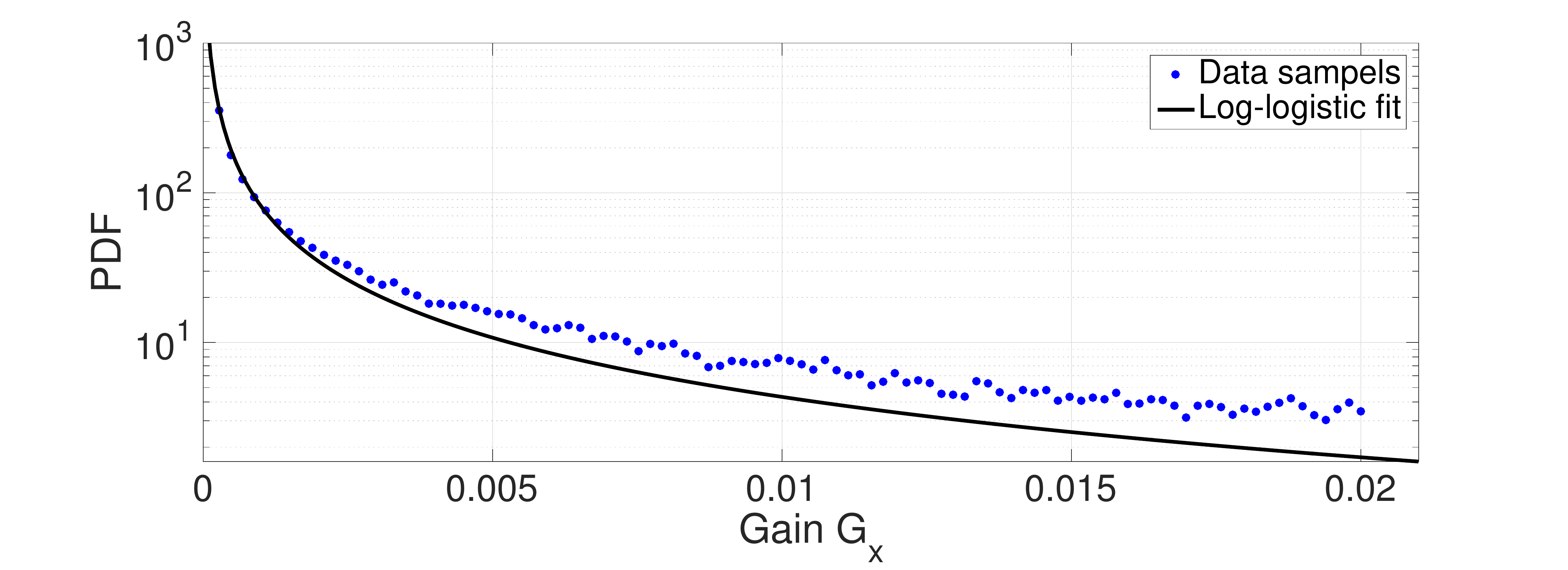}
\caption{Fitting of the aligned gain $G_x^{(\ISO)}$ with the $\ISO$ element pattern. The empirical PDF of $G_x^{(\ISO)}$ is obtained by the NYU mmWave network simulator~\cite{mezzavilla15}, and is fit with the \emph{log-logistic} distribution in Remark~3 ($n_{\text{RX}} = 64$, $n_{\text{TX}} = 256$).}
\label{log-logistic_fit_Gx}
\end{figure}

Following the same procedure as used for the aligned gain with the NYU simulator, we extract the distribution of the misaligned gain $G_x^{(z)}$ under the $\ISO$ and $\3G$ element patterns.
With the $\ISO$ element pattern, we found that the $G_x^{(\ISO)}$ PDF displayed in Fig.~\ref{log-logistic_fit_Gx} has a steep decreasing slope in the vicinity of zero, while showing a heavier tail than the exponential distribution.
This implies that the occurrence of strong interference is not frequent thanks to the sharpened mainlobe beams, yet is still non-negligible due to the interference from sidelobes that include the backward propagation. We examined possible distributions satisfying the aforementioned two characteristics, and conclude that a \emph{log-logistic} distribution provides the most accurate fitting result with the simulated misaligned gain.

\begin{remark} (Misaligned Gain, $\ISO$) \emph{At the typical UE, and using $\ISO$ antenna elements, the misaligned gain $G_x^{(\ISO)}$ can be approximated by a \emph{log-logistic} distribution with PDF
\begin{align}
f_{G_x}^{(\ISO)}(y;a,b) =\frac{\(\frac{b}{a}\)\(\frac{y}{a}\)^{b -1}}{\(1+\(\frac{y}{a}\)^b\)^2}
\end{align}
where the values of $a$ and $b$ are provided in Tab.~\ref{evolution_a_b_factors}.}
\end{remark}    

A log-logistic distribution is given by a random variable whose logarithm has a logistic distribution. The shape is similar to a log-normal distribution, but has a heavier tail~\cite{Bennett:83}. For a similar reason addressed after Remark~2, a log-logistic distribution is determined by two parameters $a$ and $b$, and their relationship with the number of antenna elements is difficult to generalize.
We instead report the appropriate values of $a$ and $b$ for 16 combinations of $n_\tx$ and $n_\rx$ in Tab.~\ref{evolution_a_b_factors}.

\begin{table}
\caption{{Misaligned gain}'s \emph{log-logistic} distribution parameters $(a,b)$ with the $\ISO$ element patterns for different $n_{\text{TX}}$ and $n_{\text{RX}}$.\vspace{\baselineskip}}
\centering
\footnotesize
\renewcommand{\arraystretch}{1.2}
\resizebox{\columnwidth}{!}{\begin{tabular}{cc||cccc}
\toprule
\multicolumn{2}{c||}{\multirow{2}{*}{$(a,b)$}}                                           & \multicolumn{4}{c}{$n_{\text{TX}}$}\\ \cline{3-6} 
\multicolumn{2}{c||}{}      & \multicolumn{1}{c|}{\textbf{4}}                                                                           & \multicolumn{1}{c|}{\textbf{16}}                                                                          & \multicolumn{1}{c|}{\textbf{64}}                                                                          & \textbf{256}                                                                         \\ \hline \hline
\multicolumn{1}{c|}{\multirow{4}{*}{$n_{\text{RX}}$}} & \textbf{4}   & \multicolumn{1}{c|}{\begin{tabular}[c]{@{}l@{}}$(3.28,0.877)$\end{tabular}} & \multicolumn{1}{c|}{\begin{tabular}[c]{@{}l@{}}$(2.51,0.743)$\end{tabular}} & \multicolumn{1}{c|}{\begin{tabular}[c]{@{}l@{}}$(2.11,0.722)$\end{tabular}} & \begin{tabular}[c]{@{}l@{}}$(1.92,0.709)$\end{tabular} \\ \cline{2-6} 
\multicolumn{1}{c|}{}                                 & \textbf{16}  & \multicolumn{1}{c|}{\begin{tabular}[c]{@{}l@{}}$-$\end{tabular}} & \multicolumn{1}{c|}{\begin{tabular}[c]{@{}l@{}}$(3.49,0.656)$\end{tabular}} & \multicolumn{1}{c|}{\begin{tabular}[c]{@{}l@{}}$(3.28,0.612)$\end{tabular}} & \begin{tabular}[c]{@{}l@{}}$(2.89,0.589)$\end{tabular} \\ \cline{2-6} 
\multicolumn{1}{c|}{}                                 & \textbf{64}  & \multicolumn{1}{c|}{\begin{tabular}[c]{@{}l@{}}$-$\end{tabular}} & \multicolumn{1}{c|}{\begin{tabular}[c]{@{}l@{}}$-$\end{tabular}} & \multicolumn{1}{c|}{\begin{tabular}[c]{@{}l@{}}$(2.55,0.57)$\end{tabular}}  & \begin{tabular}[c]{@{}l@{}}$(1.98,0.551)$\end{tabular} \\ \cline{2-6} 
\multicolumn{1}{c|}{}                                 & \textbf{256} & \multicolumn{1}{c|}{\begin{tabular}[c]{@{}l@{}}$-$\end{tabular}} & \multicolumn{1}{c|}{\begin{tabular}[c]{@{}l@{}}$-$\end{tabular}} & \multicolumn{1}{c|}{\begin{tabular}[c]{@{}l@{}}$-$\end{tabular}} & \begin{tabular}[c]{@{}l@{}}$(1.45,0.547)$
\end{tabular}
\\
\bottomrule
\end{tabular}}
\label{evolution_a_b_factors}
\end{table}

\begin{figure}[t!]
\centering
\setlength{\belowcaptionskip}{-0.4cm}
\includegraphics[width=\columnwidth]{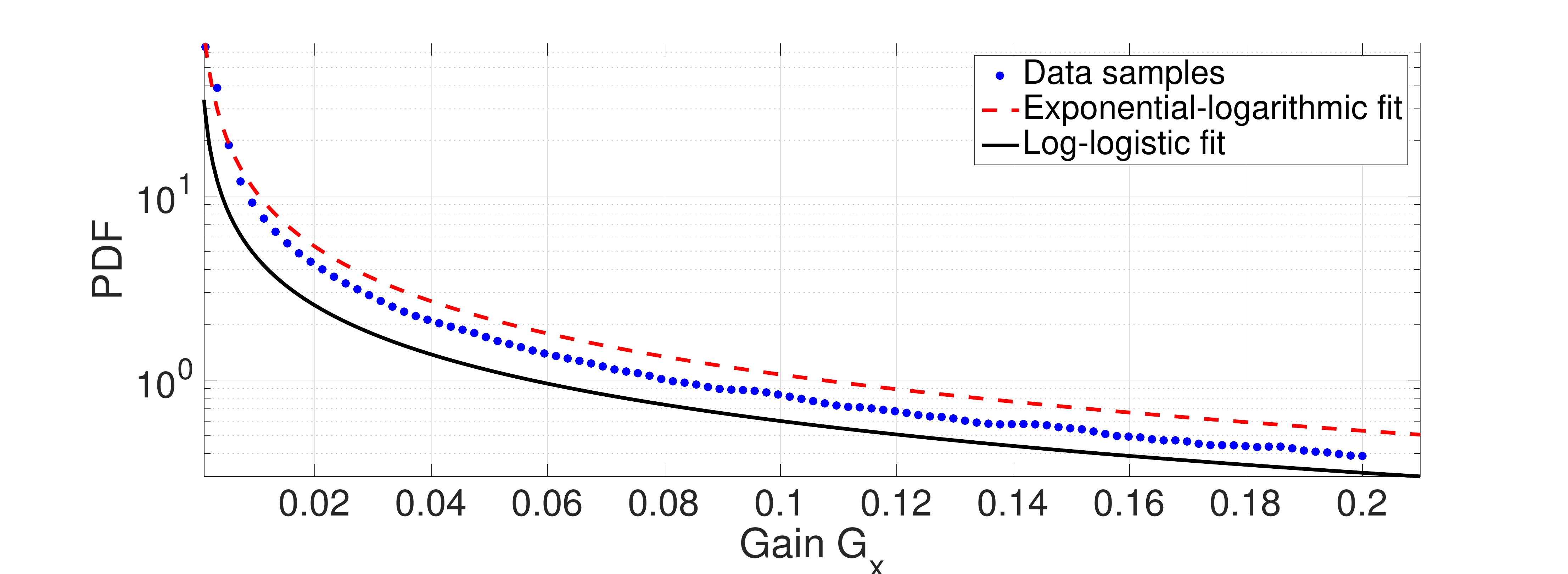}
\caption{Fitting of the aligned gain $G_x^{(\3G)}$ with the $\3G$ element pattern. The empirical PDF of $G_x^{(\3G)}$ fits with the \emph{exponential-logarithmic} distribution in Remark~4. It no longer fits with a log-logistic distribution, as opposed to the $\ISO$ element pattern's ($n_{\text{RX}} = 64$, $n_{\text{TX}} = 256$).}
\label{comparison_dist_gx}
\end{figure}

Next, with the $\3G$ element pattern, we identified the $G_x^{(\3G)}$ PDF in Fig.~\ref{comparison_dist_gx}.
Using the simulated data samples we have performed a test on the decay of the tail in order to understand if the behavior was heavy tailed.
It turns out that the PDF of $G_x^{(\3G)}$ has a lighter tail than the exponential distribution, which is far different from the heavy-tailed $G_x^{(\ISO)}$ distribution.
In this case, we found that the misaligned gain $G_x^{(\3G)}$ fits well an exponential-logarithmic distribution, as also used for the aligned gain $G_o^{(\3G)}$ in Remark~2. 
\begin{remark} (Misaligned Gain, $\3G$) \emph{At the typical UE, and adopting the $\3G$ element pattern, the misaligned gain $G_x^{(\3G)}$ can be approximated by an \emph{exponential-logarithmic} distribution with PDF
\begin{align}
f_{G_x}^{(\3G)}(y;b_x,p_x) = \frac{1}{- \ln (p_x)} \frac{b_x(1-p_xe^{-b_x y})}{1-(1-p_x)e^{-b_x y}}.
\end{align}
where the values of parameters $b_x$ and $p_x$ are provided in~Tab.~\ref{table_beta_p_x}.}
\end{remark}

Although both $G_o^{(\3G)}$ and $G_x^{(\3G)}$ can be described by using exponential-logarithmic distributions, these two results come from different reasons, respectively.
For $G_o^{(\3G)}$, it follows from the higher mainlobe gains than under the $\ISO$ element pattern that yields the exponentially distributed $G_o^{(\ISO)}$.
For $G_x^{(\3G)}$, on the contrary, its light-tailed distribution originates from attenuating sidelobes, reducing the interfering probability.
For these distinct reasons, the distribution parameters $(b_o,p_o)$ for $G_o^{(\3G)}$ and $(b_x,p_x)$ for $G_x^{(\3G)}$ are different, as shown in Tab.~\ref{table_beta_p} and Tab.~\ref{table_beta_p_x}.
Moreover, we note that in order to precisely fit both the distributions for the $\mathsf{3GPP}$ case, due to the particular behavior of both tail and slope parts we have studied several well known distributions.
We have evaluated the accuracy by measuring the root-mean-square error (RMSE) and obtained Tab.~\ref{table_rmse}.
By evaluating the RMSE, we have concluded that the exponential-logarithmic distribution was the most accurate distribution, among the ones evaluated, for both $G_o^{(\3G)}$ and $G_x^{(\3G)}$.
\begin{table}
\caption{{Misaligned gain}'s \emph{exponential-logarithmic} distribution parameters $(b_x,p_x)$ with the $\3G$ element patterns for different $n_{\text{TX}}$ and $n_{\text{RX}}$.}
\centering
\renewcommand{\arraystretch}{1.2}
\resizebox{\columnwidth}{!}{\begin{tabular}{cc||cccc}
\toprule
\multicolumn{2}{c||}{\multirow{2}{*}{$(b_x,p_x)$}}                                           & \multicolumn{4}{c}{$n_{\text{TX}}$}                                                                                                                                                                                                                                                                                                                                                  \\ \cline{3-6} 
\multicolumn{2}{c||}{}      & \multicolumn{1}{c|}{\textbf{4}}                                                                           & \multicolumn{1}{c|}{\textbf{16}}                                                                          & \multicolumn{1}{c|}{\textbf{64}}                                                                          & \textbf{256}                                                                         \\ \hline \hline
\multicolumn{1}{c|}{\multirow{4}{*}{$n_{\text{RX}}$}} & \textbf{4}   & \multicolumn{1}{c|}{\begin{tabular}[c]{@{}l@{}}$(4.428,4.3\text{e-}5)$\end{tabular}} & \multicolumn{1}{c|}{\begin{tabular}[c]{@{}l@{}}$(0.7967,3.7\text{e-}5)$\end{tabular}} & \multicolumn{1}{c|}{\begin{tabular}[c]{@{}l@{}}$(0.288,6.8\text{e-}5)$\end{tabular}} & \begin{tabular}[c]{@{}l@{}}$(1.2\text{e-04},1.5\text{e-}9)$\end{tabular} \\ \cline{2-6} 
\multicolumn{1}{c|}{}                                 & \textbf{16}  & \multicolumn{1}{c|}{\begin{tabular}[c]{@{}l@{}}$-$\end{tabular}} & \multicolumn{1}{c|}{\begin{tabular}[c]{@{}l@{}}$(0.2873,6.5\text{e-}5)$\end{tabular}} & \multicolumn{1}{c|}{\begin{tabular}[c]{@{}l@{}}$(0.024,3.6\text{e-}5)$\end{tabular}} & \begin{tabular}[c]{@{}l@{}}$(0.075,7.4\text{e-}7)$\end{tabular} \\ \cline{2-6} 
\multicolumn{1}{c|}{}                                 & \textbf{64}  & \multicolumn{1}{c|}{\begin{tabular}[c]{@{}l@{}}$-$\end{tabular}} & \multicolumn{1}{c|}{\begin{tabular}[c]{@{}l@{}}$-$\end{tabular}} & \multicolumn{1}{c|}{\begin{tabular}[c]{@{}l@{}}$( 0.2316,1.5\text{e-}4)$\end{tabular}}  & \begin{tabular}[c]{@{}l@{}}$(0.0133,2.34\text{e-}5)$\end{tabular} \\ \cline{2-6} 
\multicolumn{1}{c|}{}                                 & \textbf{256} & \multicolumn{1}{c|}{\begin{tabular}[c]{@{}l@{}}$-$\end{tabular}} & \multicolumn{1}{c|}{\begin{tabular}[c]{@{}l@{}}$-$\end{tabular}} & \multicolumn{1}{c|}{\begin{tabular}[c]{@{}l@{}}$-$\end{tabular}} & \begin{tabular}[c]{@{}l@{}}$(0.2406,2.7\text{e-}4)$
\end{tabular}
\\
\bottomrule
\end{tabular}}
\label{table_beta_p_x}
\end{table}

\begin{table}
\centering 
\footnotesize
\caption{Minimized RMSE for aligned and misaligned gains under different fitting distributions (for the case when the fitted distribution shape was unable to match the data, we marked it as avoid).}
\renewcommand{\arraystretch}{1}
\begin{tabular}{r c c} 
\toprule
\multirow{2}{*}{Distribution Type}& \multicolumn{2}{c}{Minimized RMSE}\\& $G_o$    & $G_x$ \\ \cmidrule(r){1-1} \cmidrule(r){2-2} \cmidrule(r){3-3}
Exponential                                                        & $1.99\text{e-}6$ & $7.46$ \\ 
Exponential-logarithmic  & $\mathbf{4.11\textbf{e-}7}$ & $\mathbf{0.51}$ \\ 
Burr                                                               & $4.26\text{e-}6$ & $1.74$ \\ 
Log-logistic                                                       & $-$       & $1.63$ \\ 
Log-normal                                                         & $-$       & $-$   \\
Log-Cauchy                                                         & $-$       & $0.56$  \\
Gamma                                                       & $-$       & $0.80$   \\
Weibull                                                         & $4.27\text{e-}6$       & $0.63$   \\
Rayleigh                                                         & $-$       & $-$   \\
Nakagami                                                        & $-$       & $1.04$   \\
L\'evi                                                        & $-$       & $1.73$   \\\bottomrule
\end{tabular}
\label{table_rmse}
\end{table}

The fitting plots of both aligned and misaligned gains, respectively Figs.~\ref{exponential_fit_G0}--\ref{comparison_dist_g0} and Figs.~\ref{log-logistic_fit_Gx}--\ref{comparison_dist_gx}, permit to see the approximation error which is introduced due to the fitting procedure.
However, we note that we are plotting the curves using a log-scale for the y-axis, thus when the PDF becomes smaller even if the error gap looks bigger, the real error may be smaller.

Note that $G_x^{(\ISO)}$ is often considered as a \emph{Nakagami-$m$} or a \emph{log-normal} distributed random variable~\cite{bai15,Andrews:17,direnzo2015}. In Sect.~\ref{numerical_results}, we will thus compare our proposed distributions for $G_x^{(z)}$ with them.
For a fair comparison, for a Nakagami-$m$ distribution with $n_\tx = 256$ and $n_\rx = 64$, we will use its best-fit distribution parameters obtained by curve-fitting with the system-level simulation, which are given with the PDF as follows.

\vspace{-10pt}\small\begin{align}
\label{Nakagami}
f_{G_x}^{(\ISO)}(y;m,g) = \frac{2m^m}{\Gamma(m) g^m} y^{2m -1} \exp\(-\frac{m}{g}y^2\), _{} \begin{cases} m = 0.099 \\ g = 50.53 \end{cases}
\end{align}\normalsize
With this PDF, we will observe in Sect.~\ref{numerical_results} that a Nakagami-$m$ distribution underestimates the tail behavior of $G_x^{(\ISO)}$ too much, thereby leading to a loose empirical upper bound for the $\SINR$ coverage probability.

Likewise, for a log-normal distribution with $n_\tx = 256$ and $n_\rx = 64$, we will consider the following PDF with the parameters.

\vspace{-10pt}\small\begin{align}
\label{log-normal}
\hspace{-10pt}f_{G_x}^{(\ISO)}(y;\sigma,\mu) &= \tfrac{1}{y\sigma \sqrt{2 \pi}}\exp\left(-\tfrac{\left(\log y-\mu\right)^2}{2 \sigma^2}\right), &\begin{cases} \sigma = 2.962 \\ \mu = 0.908 \end{cases}
\end{align}\normalsize
Under the $\ISO$ element pattern, it will be shown in Sect.~\ref{numerical_results} that a log-normal distribution is a better fit than a Nakagami-$m$ distribution, yet it still underestimates the interference, yielding an empirical upper bound to the $\SINR$ coverage probability. 

As an auxiliary result, we will also provide the result with a \emph{Burr} distribution~\cite{burr1942}.
This overestimates the tail behavior of $G_x^{(\ISO)}$, leading to the empirical lower bound of the $\SINR$ coverage probability. For this, we will consider the following PDF under $n_\tx = 256$ and $n_\rx = 64$. 

\vspace{-10pt}\small\begin{align}
\label{burr}
f_{G_x}^{(\ISO)}(y;c,k) &= \tfrac{c k y^{c-1}}{(1+y^c)^{k+1}},  &\begin{cases} c = 0.692 \\ k = 0.518 \end{cases}
\end{align}\normalsize

\section{mmWave $\SINR$ Coverage Probability} 
\label{sinr_coverage}

In this section, we aim at deriving the closed-form expression of the $\SINR$ coverage probability $\C(T)$, defined as the probability that the typical UE's $\SINR$ is no smaller than a target $\SINR$ threshold $T>0$, i.e., $\C(T):=\Pr(\SINR \geq T)$. In the first subsection, utilizing the aligned/misaligned gains provided in Sect.~III, we derive the exact $\SINR$ coverage expressions under $\ISO$ and $\3G$ element patterns. In the following subsection, applying a first-moment approximation to aligned/misaligned gains, we further simplify the $\SINR$ coverage expressions.

\setcounter{equation}{27}
\begin{figure*}[b]
\rule[0.5ex]{\linewidth}{1pt}
\footnotesize\begin{align}
\label{la_place_functional}
\mathcal{L}_{I^j_i}(r) &=  \exp \left( -2 \pi \lambda_b \int_0^{\infty} \left( \int_{\left(\frac{\beta_j r^{\alpha_i}}{\beta_i} \right)^{\frac{1}{\alpha_j}}}^{\infty} \left[1- \exp\left(-   \frac{\ell^j(v)\mu_o T g }{\ell^i(r)} \right)\right] v p_j(v) \mathrm{d}v \right)  f_{G_x}^{(z)}(g) \mathrm{d}g \right) 
\end{align}\normalsize
\end{figure*}
\setcounter{equation}{22}

\subsection{$\SINR$ Coverage}
Let $r_{x_o}^i$ denote the association distance of the typical UE associating with $x_o\in\Phi_i$. By using the law of total probability, $\C$ at the typical UE can be represented as

\vspace{-10pt}\small\begin{align}
&\C (T) = \Pr \big( \underbrace{\SINR \geq T,  x_o\in \Phi_L}_{\SINR_L\geq T} \big) + \Pr\big( \underbrace{\SINR \geq T,  x_o\in \Phi_N}_{\SINR_N \geq T} \big)\\
&= \E_{r_{x_0}^L}\left[\Pr \big( \SINR_L\geq T | r_{x_o}^L \big)\right] + \E_{r_{x_o}^N}\left[\Pr\big( \SINR_N \geq T |r_{x_o}^N \big)\right].
\label{sinr_coverage_definition}
\end{align}\normalsize
In \eqref{sinr_coverage_definition}, two expectations are taken over the typical UE's association distance $r_{x_o}^i$. The PDF of $r_{x_o}^i$ is given by~\cite{Gupta2016} as
\small\begin{align}
&f_{r_{x_o}^i}(r) := f_{|x_o|,i}(r, x_o\in\Phi_i)\\
&= 2 \pi \lambda_i(r) r\exp \Bigg(- 2    \pi \lambda_b \bigg[ \int_0^{r} v p_i(v)\mathrm{d}v + \int_0^{ \left({r^{\alpha_i} \beta_{i^\prime} /\beta_{i}}\right)^{\frac{1}{\alpha_{i^\prime}}} } \hspace{-30pt} v p_{i'}(v)\mathrm{d}v \bigg] \Bigg)  \label{Eq:DistPDF}
\end{align}\normalsize
where $\lambda_i(r)=\lambda_b p_i(r)$, and $i^\prime$ indicates the opposite LoS/NLoS state with respect to $i$.

For the $\ISO$ element pattern, the typical UE's $\SINR$ coverage probability $\C(T)$ in \eqref{sinr_coverage_definition} is then derived by exploiting $f_{r_{x_o}^i}(r)$ while applying Campbell's theorem~\cite{HaenggiSG} and the $G_o^{(\ISO)}$ distribution in Remark 1.
\begin{proposition}\linespread{1.2} (Coverage, $\ISO$) \emph{At the typical UE, and considering arrays with $\ISO$ radiation elements, the $\SINR$ coverage probability $\C(T)$ for a target $\SINR$ threshold $T>0$ is given as
\small\begin{align}
\C(T) &= \sum_{i\in \{L,N\}} \int_0^{\infty} f_{r_{x_o^i}} \left(r\right) 
\exp \left(\frac{ -\mu_o T r^{\alpha_i} \sigma^2 }{\beta_i} \right)
\nn\\
&\hspace{40pt}\times \mathcal{L}_{I_i^L}\left( \frac{\mu_o T}{ \ell^i(r) }\right) \mathcal{L}_{I_i^N}\left(\frac{\mu_o T}{\ell^i(r) }\right) \mathrm{d}r,
\end{align}\normalsize
where $\mathcal{L}_{I_i^j}(r)$ is the Laplace transform of the interference from BSs $\in \phi_j$, for $j\in\{L,\,N\}$, to the typical UE and is given in~\eqref{la_place_functional} with $z=\ISO$.
}

\noindent Sketch of the Proof: \emph{Starting from the $\SINR$ joint probability in~\eqref{sinr_coverage_definition} and applying the $\SINR$ definition we obtain an expression which depends on the CCDF $F_{G_o}^{(\ISO)}(y; \mu_o)$.
Then, applying Remark 1, which provides a channel gain expression with specific distribution, together with Slyvnyak's theorem and the mutual independence of PPPs $\Phi_i^L$ and $\Phi_i^N$ we obtain the final coverage expression.
The detailed proof is provided in Appendix I.}\hfill $\blacksquare$
\end{proposition}
\setcounter{equation}{28}

\noindent Note that $1/\mu_o$ is the mean aligned gain in Remark 1.
The misaligned gain PDF $f_{G_x}^{(\ISO)}(y)$ and its corresponding parameters are provided in Remarks 3 and 4 as well as in Tab.~\ref{evolution_a_b_factors}.
As opposed to the standard method where the exponential random variables can be found in both desired and interfering links, the misalignment gain in our interfering link follows a log-logistic distribution.
This does not allow to further expand the expression as done in the standard method, yet the expression can easily be calculated numerically as done in~\cite{Haenggi18}, which is far simpler than the system-level simulation complexity.
Then, the term $p_i$ is the LoS/NLoS channel state probability defined in Sect.~\ref{propagation_and_channel_gain}.

For the $\3G$ element pattern, following the same procedure and $G_o^{(\3G)}$ distribution in Remark 2, we obtain $\C(T)$ as shown in the following proposition.

\begin{proposition}\linespread{1.2} (Coverage, $\3G$) \emph{At the typical UE, and considering arrays with $\3G$ radiation elements, the $\SINR$ coverage probability $\C(T)$ for a target $\SINR$ threshold $T>0$ is upper bounded as}
\emph{\small\begin{align}
&\C(T) \le \sum_{i\in \{L,N\}} \int_0^{\infty} \frac{f_{r_{x_o^i}} \left(r\right)}{\ln \left(p_o\right) }
\ln \bigg(1- \left(1-p_o \right)\nn\\
&\times \exp \left(\frac{ -b_o T r^{\alpha_i} \sigma^2 }{\beta_i} \right)
\mathcal{L}_{I_i^L}\left( \frac{b_o T}{ \ell^i(r) }\right) \mathcal{L}_{I_i^N}\left(\frac{b_o T}{\ell^i(r) }\right)\bigg)\mathrm{d}r,
\end{align}\normalsize
where the Laplace transform $\mathcal{L}_{I_i^j}(r)$ for $j\in\{L,N\}$ is given in \eqref{la_place_functional} with $z=\3G$ at the bottom of this page.}

\noindent Sketch of the Proof: \emph{The first step of the demonstration is equivalent to the one in Proposition 1 with the only difference that $G_o^{(\3G)}$ follows an exponential-logarithmic distribution with the CCDF $F(y;b_o,p_o) =  {\ln \left( 1-\left(1-p_o\right) e^{-b_o y} \right)}/{\ln p_o}$. Then, differently from the previous proposition, Jensen's inequality is used to obtain an upper bound of the $\SINR$ coverage probability. The remainder of the proof follows the Proof of Proposition~1. For completeness, the detailed derivation is provided in Appendix~II.}\hfill $\blacksquare$
\end{proposition}

It is worth noting that the Laplace transform expression in~\eqref{la_place_functional} is used for both $\ISO$ and $\3G$ element patterns, i.e., in Propositions 1 and 2. Here, the element pattern is differentiated only by the distribution of the misaligned gain $f_{G_x}^{(z)}(g)$ contained therein. For different element patterns and their fitting results, we can thus change $f_{G_x}^{(z)}(g)$ accordingly while keeping the rest of the terms, thereby allowing us to quickly compare the resulting $\SINR$s.
This is an advantage of the analysis, that avoids redundant calculations. 



\setcounter{equation}{36}
\begin{figure*}[b]
\rule[0.5ex]{\linewidth}{1pt}
\footnotesize\begin{align}
\label{la_place_functional_hat}
\hat{\mathcal{L}}_{I_i^j}(s) &= \exp \Bigg( -2 \pi \lambda_b \int_{ \left( \frac{\beta_{j}}{\beta_i} r^{\alpha_{i}} \right)^{\frac{1}{\alpha_{j}}}}^\infty
\Bigg[ \frac{\varphi_\tx\varphi_\rx}{ 4\pi^2} F\left(M_\tx^{(z)} M_\rx^{(z)}\right)
+\frac{\varphi_\tx}{ 2\pi} \left(1 - \frac{\varphi_\rx}{ 2\pi}\right) F\left(M^{(z)}_\tx m^{(z)}_\rx \right) \nn\\
&\quad\quad\quad\quad + \left(1 - \frac{\varphi_\tx}{ 2\pi}\right)\frac{\varphi_\rx}{ 2\pi} F\left(m^{(z)}_\tx M^{(z)}_\rx \right)
+ \left(1 - \frac{\varphi_\tx}{ 2\pi}\right) \left(1 - \frac{\varphi_\rx}{ 2\pi}\right) F\left(m^{(z)}_\tx m^{(z)}_\rx \right)
\Bigg] v p_j(v) \mathrm{d}v  \Bigg)
\end{align}
where $F\left(x\right) = {sx v^{-\alpha_i}\beta_i}/({1+sx v^{-\alpha_i}\beta_i})$.
\end{figure*}
\setcounter{equation}{29}

\subsection{Simplified $\SINR$ coverage} \label{Sect:simplified}
In this subsection, our goal is to further simplify the $\SINR$ coverage probability expressions in Propositions 1 and 2.
To this end, we revisit a channel-antenna gain approximation approach that is commonly used with stochastic geometric analysis, as done in~\cite{bai15,direnzo2015,Andrews:17,park2016,Gupta2016,li16,Kim:18}.
This approach relies on approximating the channel gain based on its first-moment value, and may therefore be less accurate compared to the simulated result. 

Nevertheless, with a slight refinement, we conjecture that such a simple approach can still provide a tight approximation, also for the $\3G$ element pattern.
In fact, the only major difference, with respect to the $\ISO$ case is the presence of a high front-back ratio, which in turn is due to the directivity gain considered.
With this purpose in mind, we elaborate the approximation procedures of the channel and antenna gains as follows.
For the channel gain, instead of directly using the realistic channel model, we consider a first-order approximated Rayleigh fading channel with the mean value that is identically set as that of the realistic channel model.
For the antenna gain, as illustrated in Fig.~\ref{example_array_radiation_pattern}, we approximate the continuous array gain using only two constants, i.e., mainlobe gain $M_s^{(z)}$ and sidelobe gain $m_s^{(z)}$. The mean aligned gain $\Upsilon_o^{(z)}$ and the mean misaligned gain $\Upsilon_x^{(z)}$ are determined by these two antenna gain constants that are specified by the $\ISO$ and $\3G$ element patterns, as detailed in the following remark.
\begin{remark} (Simplified Aligned/Misaligned Gains) \emph{For a given antenna array radiation pattern $z\in\{\ISO, \3G\}$, we consider the following channel and array radiation approximations.
\begin{itemize}
\item \emph{Rayleigh fading channel gain} -- Both the aligned gain $G_o^{(z)}$ and the misaligned gain $G_x^{(z)}$ at the typical UE independently follow an \emph{exponential} distribution, i.e., 
\begin{align}
\hspace{-5pt} G_o^{(z)}\sim\textsf{Exp}(1/\Upsilon_o^{(z)}) \quad \text{and}\quad  G_x^{(z)}\sim\textsf{Exp}(1/\Upsilon_x^{(z)}).
\end{align}
\item \emph{Piece-wise constant array gain} -- The average channel gains $\Upsilon_o^{(z)}$ and $\Upsilon_x^{(z)}$, taken from~\cite{bai15}, are given as:
\begin{align}
\Upsilon_o^{(z)} &= \;\; \; M_\tx^{(z)}M_\rx^{(z)}\quad \text{and}\\
\Upsilon_x^{(z)} &=
\begin{cases}
M_\tx^{(z)} M_\rx^{(z)} \quad &\text{w.p.} \quad  \frac{\varphi_{\tx}}{2 \pi} \frac{\varphi_{\rx}}{2 \pi}\\
M_\tx^{(z)}m_\rx^{(z)} \quad &\text{w.p.}\quad  \frac{\varphi_{\tx}}{2 \pi}(1 - \frac{\varphi_{\rx}}{2 \pi})\\
m_\tx^{(z)} M_\rx^{(z)} \quad &\text{w.p.} \quad (1 -\frac{\varphi_{\tx}}{2 \pi}) \frac{\varphi_{\rx}}{2 \pi}\\
m_\tx^{(z)} m_\rx^{(z)} \quad &\text{w.p.}\quad (1- \frac{\varphi_{\tx}}{2 \pi})(1- \frac{\varphi_{\rx}}{2 \pi}),
\end{cases}
\label{gi_expression}
\end{align}
where the mainlobe gain $M_s^{(z)}$ and the sidelobe gain $m_s^{(z)}$ are set as
\begin{align}
M_s^{(\ISO)} &= n_{s}  \\
M_s^{(\3G)} &= 10^{0.8} n_{s} \label{equations_model_1a_main}\\
m_s^{(z)} &= {1}/{\sin^2\left(\frac{3 \pi}{2\sqrt{n_s}}\right)},
\label{equations_model_1a_side}
\end{align}
and $n_s$ with $s\in\{\tx,\rx\}$ denotes the number of the transmit/receive antenna elements.
\end{itemize}}
\end{remark}

With the $\ISO$ element pattern, it is noted that the said simplified model becomes identical to the model considered in~\cite{Gupta2016}. In this case, the sidelobe gain $m_s^{(z)}$ in \eqref{equations_model_1a_side} is obtained from the array's 3~dB beamwidth\footnotemark~that equals~${\sqrt{3/n_s}}$.

\footnotetext{Note that the previously defined $\theta_{3\text{dB}}$ and $\phi_{3\text{dB}}$ parameters were determined by the 3~dB beamwidth of the element radiation pattern, whereas $\varphi_{s}$ is given by the 3~dB beamwidth of the array radiation pattern.} 

With the $\3G$ element pattern, by constrast, the mainlobe gain in \eqref{equations_model_1a_main} is $10^{0.8}\approx 6.31$ times higher than in the $\ISO$ radiation case, due to its maximum $8$ dBi directivity gain at each antenna element as discussed in Section \ref{antenna_gain}.
The sidelobe gain in \eqref{equations_model_1a_side} is computed in the same manner for both $\ISO$ and $\3G$ element patterns, yet has the different physical meanings for each case as detailed next.

\begin{figure}[t!]
\centering
\setlength{\belowcaptionskip}{-0.4cm}
\includegraphics[width=\columnwidth]{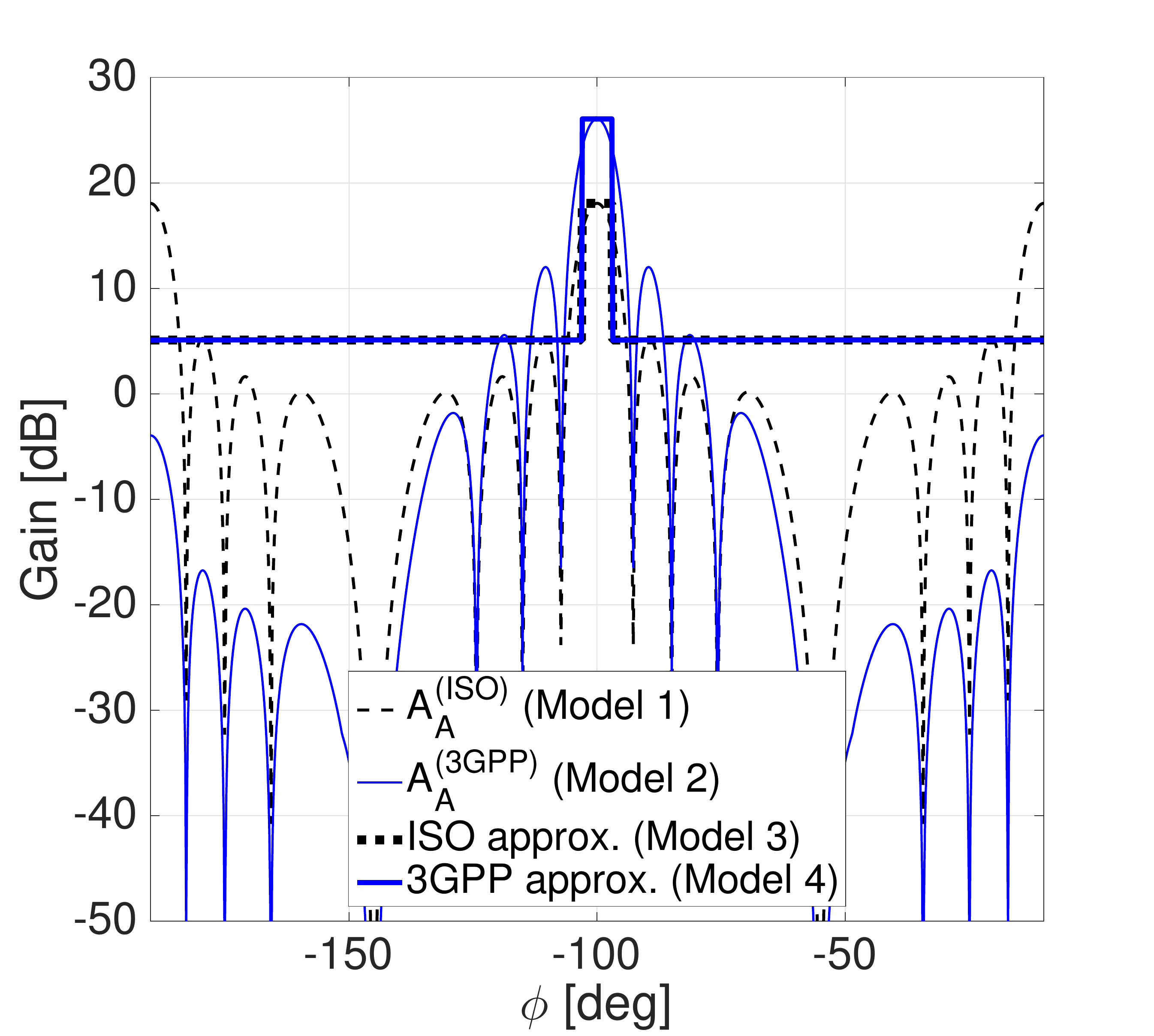}
\caption{Comparison between the array radiation gains with the $\ISO$ and $\3G$ element patterns and their piece-wise constant approximated gains given in Remark~5, with respect to the horizontal steering angle $\phi\in[-180^{\circ},180^{\circ}]$ while the vertical steering angle $\theta$ is fixed at $90^\circ$.
}
\label{example_array_radiation_pattern_complete}
\end{figure}

Following~\cite{Gupta2016}, the sidelobe gain in \eqref{equations_model_1a_side} with the $\ISO$ element pattern corresponds to the \emph{second maximum} lobe gain, as shown in Fig.~\ref{example_array_radiation_pattern_complete}. On the contrary, \eqref{equations_model_1a_side} with the $\3G$ element pattern is mostly below the second maximum lobe gain. This implicitly captures the $\3G$ element pattern's sidelobe reduction as shown in Fig.~\ref{example_array_radiation_pattern_complete}. 

Unlike the $\ISO$ element pattern, it is noted that \eqref{equations_model_1a_side} with the $\3G$ element pattern approximates the \emph{third maximum} lobe gain on average, but is not always identical to the third maximum value. In fact, due to the element-wise beam steering, the antenna gain under the $\3G$ element pattern is not symmetrical about the steering angle, so each lobe's gain can only be ordered for a given steering angle, as further explained in~\cite{rebato18}.

\begin{table*}[]
\centering
\renewcommand{\arraystretch}{1.5}
\caption{List of the channel-antenna configurations considered in Sect.~\ref{numerical_results}.}
\label{table_all_models}
\footnotesize\begin{tabular}{l c c l}
\toprule
 Configuration & \textbf{Channel} & \textbf{Antenna element radiation} & \textbf{Array radiation} \\ \cmidrule(r){1-1} \cmidrule(r){2-2} \cmidrule(r){3-3}\cmidrule(r){4-4}

\textbf{Model~1}~\cite{rebato17} & NYU~\cite{akdeniz14} & $\ISO$ & continuous main/sidelobes\\ 
\textbf{Model~2} & NYU~\cite{akdeniz14} & $\3G$~\cite{mmWave_3gpp_channel} & continuous main/sidelobes with smaller sidelobe radiations \\ 
\textbf{Model~3}~\cite{Gupta2016} & Rayleigh & $-$ & piece-wise constant main/sidelobes ($M^{(\ISO)}$ or $m^{(\ISO)}$)\\ 
\textbf{Model~4} & Rayleigh & $-$ & piece-wise constant main/sidelobes ($M^{(\3G)}$ or $m^{(\3G)}$)
\\ 
\bottomrule      
\end{tabular}
\end{table*}

Finally, utilizing the aligned and misaligned gains in Remark~5, we obtain the simplified $\SINR$ coverage probability.
\begin{proposition}\linespread{1.2} (Simplified Coverage) \emph{Using the simplified aligned and misaligned gains in Remark~5, the simplified $\SINR$ coverage probability $\hC(T)$ at the typical UE with a target SINR threshold $T>0$ is given~by}
\small\begin{align}
\hC(T) &= \sum_{i\in\{L,\,N\}} \int_0^\infty  f_{r_{x_o^i}} \left(r\right)
\exp \left( - \frac{  T r^{\alpha_i}\sigma^2}{\beta_i M_\tx^{(z)} M_\rx^{(z)}} \right)\nn\\
&\quad\; \times 
\hat{\mathcal{L}}_{I_i^L}\left(  \frac{ T (\ell^i(r))^{-1}}{M_\tx^{(z)}M_\rx^{(z)}} \right) 
\hat{\mathcal{L}}_{I_i^N}\left(  \frac{ T (\ell^i(r))^{-1}}{M_\tx^{(z)}M_\rx^{(z)}} \right) \mathrm{d}r,
\end{align}\normalsize
where $\hat{\mathcal{L}}_{I_i^j}(t)$ is given at the bottom of this page.\\
\noindent Proof: \emph{See Theorem 1 in~\cite{Gupta2016}.}\hfill $\blacksquare$
\end{proposition}

\noindent In the next section, we will validate that this simplified $\SINR$ coverage expression becomes accurate for the $\3G$ element pattern, as conjectured at the beginning of this subsection. 


\section{Numerical Results and Comparisons}
\label{numerical_results}
In this section, by using the NYU mmWave network simulator~\cite{mezzavilla15}, we validate our analytical mmWave $\SINR$ coverage expressions with the ISO element pattern in Proposition~1 and the expression with the $\3G$ element pattern in Proposition~2, as well as their simplified $\SINR$ coverage expressions proposed in~Proposition~3.
For easier comparison, the channel-antenna configurations considered in this section are categorized as four models as summarized in Tab.~\ref{table_all_models}.
The antenna configurations are illustrated in Fig.~\ref{example_array_radiation_pattern_complete}, and the channel configurations are detailed in Sect. II-B and Remark 5.
Other simulation parameters are: carrier frequency $f=28$~GHz, bandwidth $W=500$~MHz, BS density $\lambda_b=100$~BSs/km$^2$ and transmission power $P_{\text{TX}}=30$~dBm.

\begin{figure}
\centering
\includegraphics[width=\columnwidth]{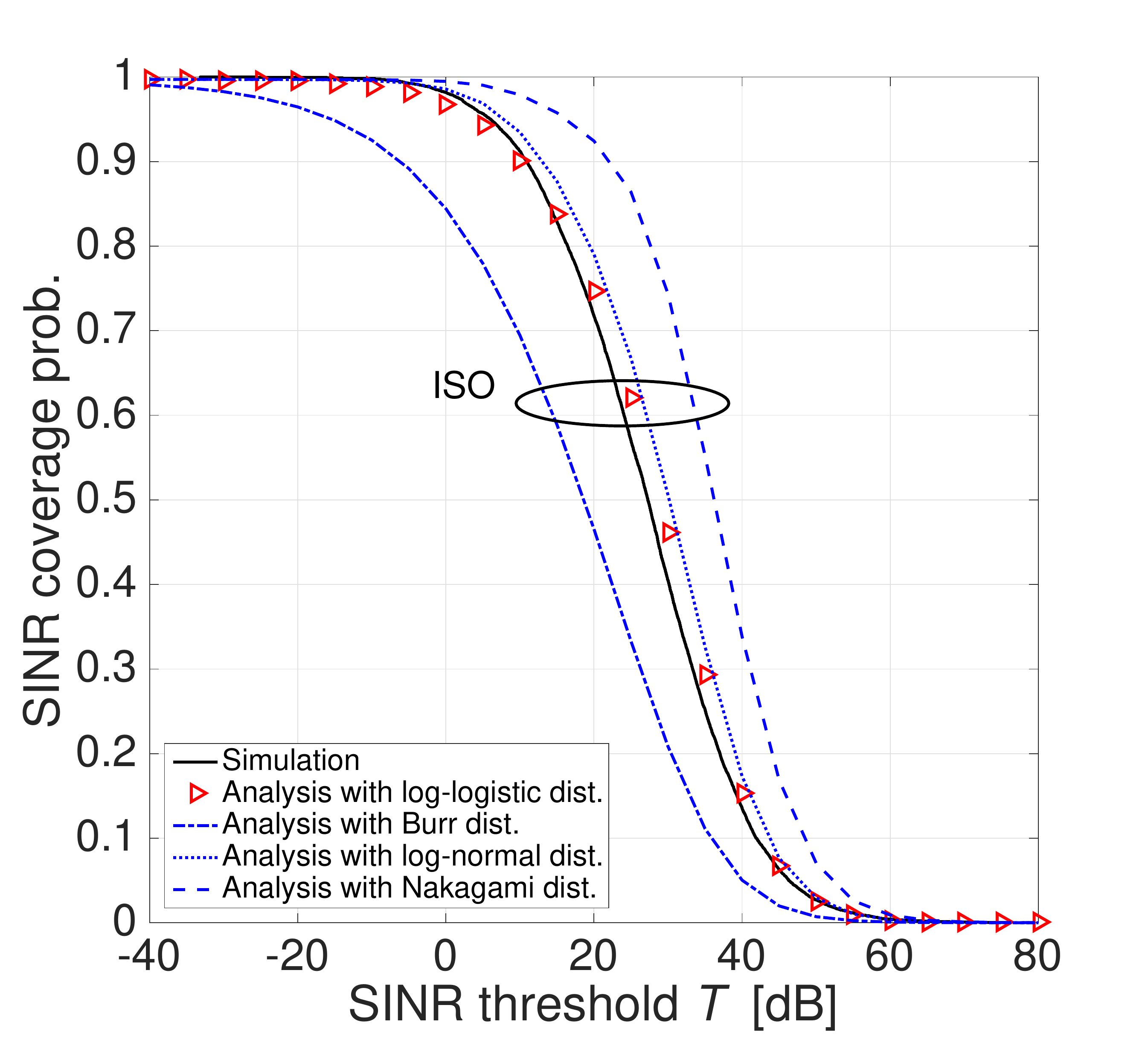} 
\caption{$\SINR$ coverage probability with the $\ISO$ element pattern under Model~1 for different misaligned gain fitting distributions: (i) the \emph{log-logistic} distribution in Remark~3, (ii) the \emph{Nakagami-m} distribution in \eqref{Nakagami}, (iii) the \emph{log-normal} distribution in \eqref{log-normal}, and (iv) the \emph{Burr} distribution in \eqref{burr}. The aligned gain is fitted with the \emph{exponential} distribution in Remark~1, and $\{n_\tx,n_\rx\}=\{256,64\}$.
}\label{bounds}
\end{figure}

\begin{figure}
\includegraphics[width=.985\columnwidth]{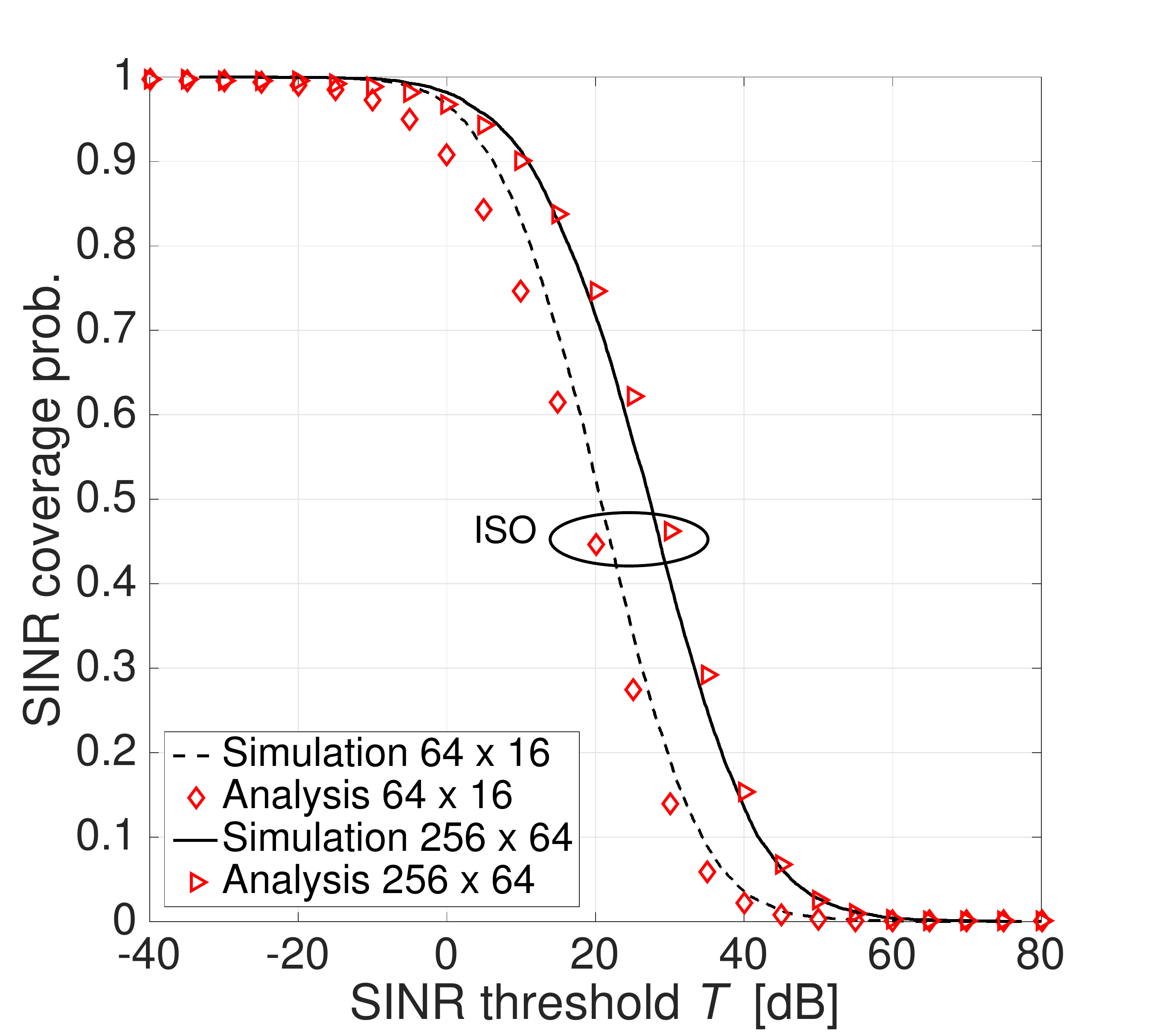}
\caption{$\SINR$ coverage probability with the $\ISO$ element pattern under Model~1. The aligned gain is fit with the \emph{exponential} distribution in Remark~1, and the misaligned gain is fitted with the \emph{log-logistic} distribution in Remark~3, for $\{n_{\text{TX}},n_{\text{RX}}\}=\{64, 16\}$ and $\{256, 64\}$.}
\label{sinr_ccdf_iso}
\end{figure}

Figs.~\ref{bounds} and \ref{sinr_ccdf_iso} show the $\SINR$ coverage probability with the $\ISO$ element pattern under Model~1.
In Fig.~\ref{bounds}, the coverage probability obtained from the NYU network simulator fits well our proposed coverage expression in Proposition~1 that utilizes the aligned gain's \emph{exponential} distribution in Remark~1 and the misaligned gain's \emph{log-logistic} distribution in Remark~3.
The proposed $\SINR$ coverage probability expression is also compared to the $\SINR$ coverage probabilities with the misaligned gain's \emph{Nakagami-$m$} and \emph{log-normal} distributions that are commonly used in stochastic geometric mmWave $\SINR$ coverage analysis~\cite{bai15,direnzo2015,Andrews:17}.
It shows that both Nakagami-$m$ and log-normal distributions given respectively in \eqref{Nakagami} and \eqref{log-normal} underestimate the interference tail behaviors, therefore yielding empirical upper bounds for the $\SINR$ coverage probability. Another misaligned gain's \emph{Burr} distribution given in \eqref{burr} by contrast yields an empirical lower bound for the $\SINR$ coverage probability. All these bounds are too loose to approximate the simulated $\SINR$ coverage probability, emphasizing our appropriate choice of the misaligned gain's log-logistic distribution.

Fig.~\ref{sinr_ccdf_iso}, by comparing the curves with the antenna element configuration $\{n_\tx,n_\rx\}=\{64,16\}$ and the curves with $\{n_\tx,n_\rx\}=\{256,64\}$, shows that the increase in the number of antenna elements not only yields a higher $\SINR$ but also makes the $\SINR$ coverage probability expression in Proposition~1 more accurate.
The latter is because the front-back ratio increases with the number of antenna elements~\cite{tse_book,rebato18}. Following a similar reasoning as discussed after Remarks~2 and 4, this reduces the impact of the high-order statistics on the alignment and misaligned gains, and thereby Proposition~1 becomes more accurate.

\begin{figure}
\centering
\setlength{\belowcaptionskip}{-0.4cm}
\centering
\hspace{-15pt}\includegraphics[width=0.98\columnwidth]{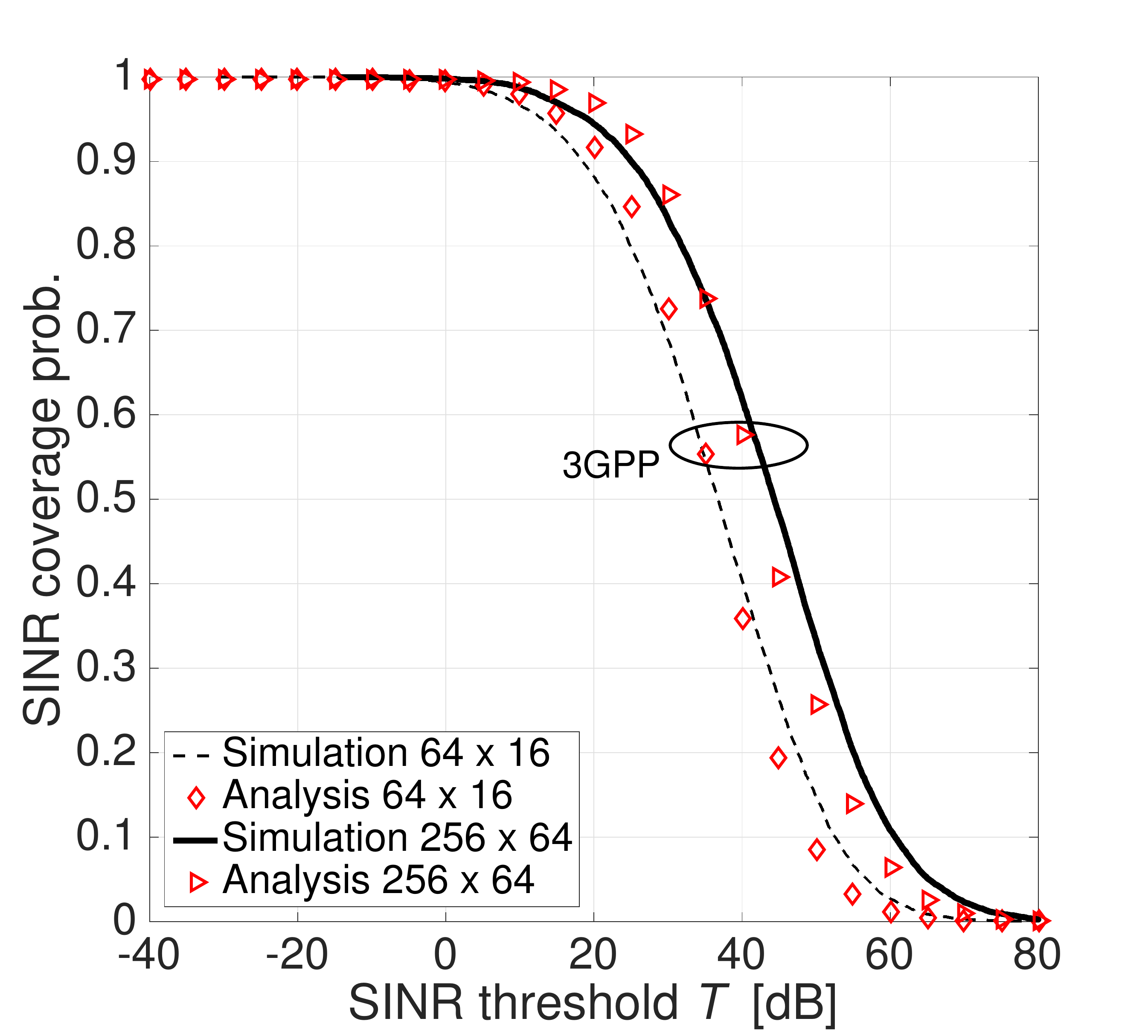}
\caption{$\SINR$ coverage probability with the $\3G$ element pattern under Model~2. The aligned and misaligned gains are fit independently with the \emph{exponential-logarithmic} distributions in Remarks~2 and 4, respectively, for $\{n_{\text{TX}},n_{\text{RX}}\}=\{64, 16\}$ and $\{256, 64\}$.}
\label{sinr_ccdf_3gpp}
\end{figure}

Next, Fig.~\ref{sinr_ccdf_3gpp} illustrates the $\SINR$ coverage probability with the $\3G$ element pattern under Model~2.
We observe that the simulated coverage probability fits well with our proposed coverage expression in Proposition~1 that utilizes the \emph{exponential-logarithmic distributions} of aligned and misaligned gains in Remarks~2 and 4, respectively.
As seen by comparing Fig.~\ref{sinr_ccdf_3gpp} to Fig.~\ref{sinr_ccdf_iso}, the $\SINR$ coverage probability with the $\3G$ element pattern is higher than the coverage probability with the $\ISO$ element pattern.
This is because of the $\3G$ element pattern's higher front-back ratio that provides higher directivity, thereby increasing the aligned gain.
It also provides lower interference that decreases the misaligned gain, consequently yielding a higher $\SINR$.
These results highlight the presence of different performance trends as the network's density increases.
This means that it is possible to accurately identify an optimal deployment density of the BSs. 
We have further studied this aspect in~\cite{rebato18}.

\begin{figure*}[h!]
\centering
\begin{subfigure}[b]{0.48\textwidth}
\centering
\includegraphics[width=\columnwidth]{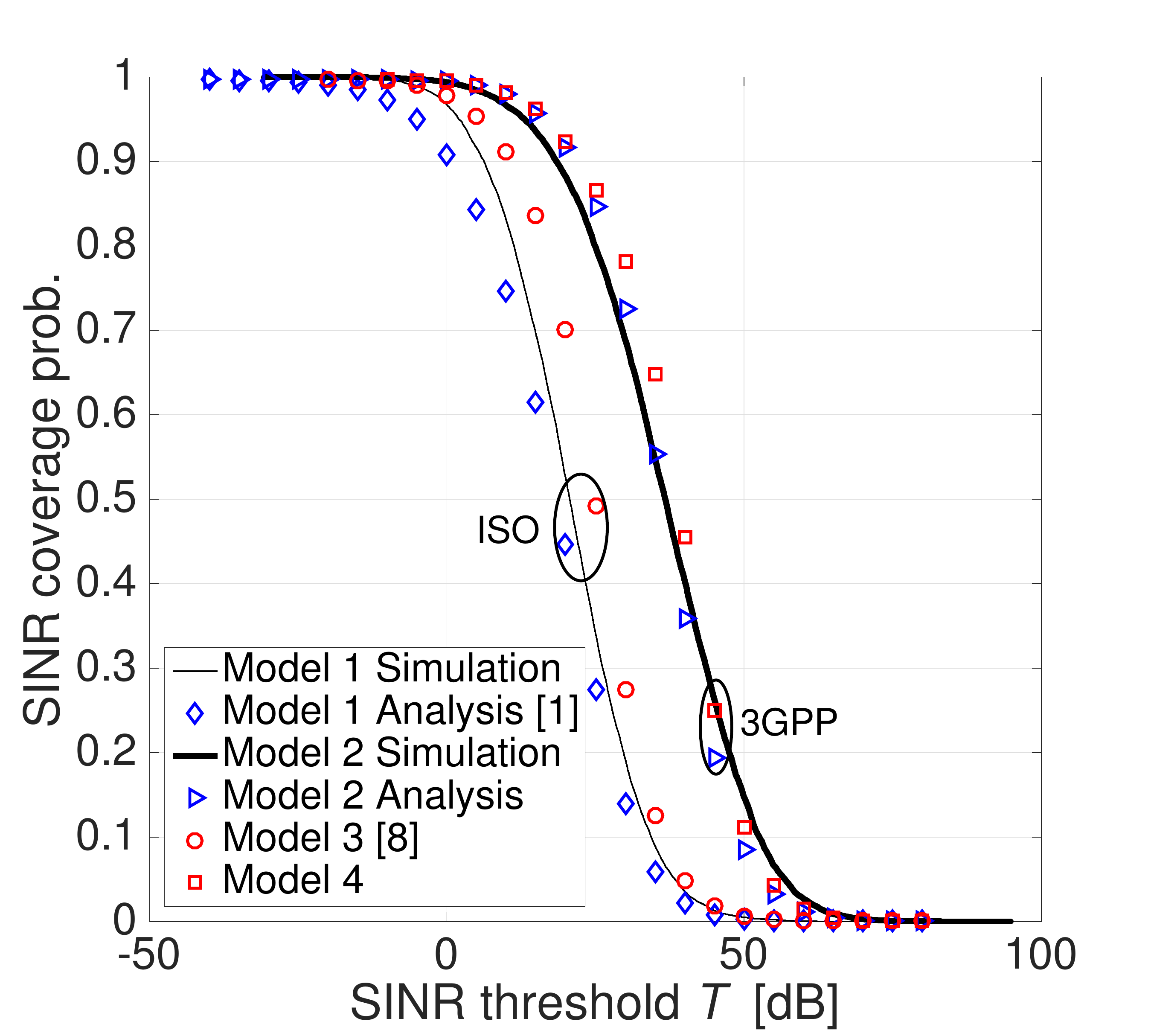}
\caption{For $\{n_{\text{TX}},n_{\text{RX}}\}=\{64, 16\}$.}
\label{comparison_64}
\end{subfigure}
\hfill
\begin{subfigure}[b]{0.48\textwidth}
\centering
\includegraphics[width=\columnwidth]{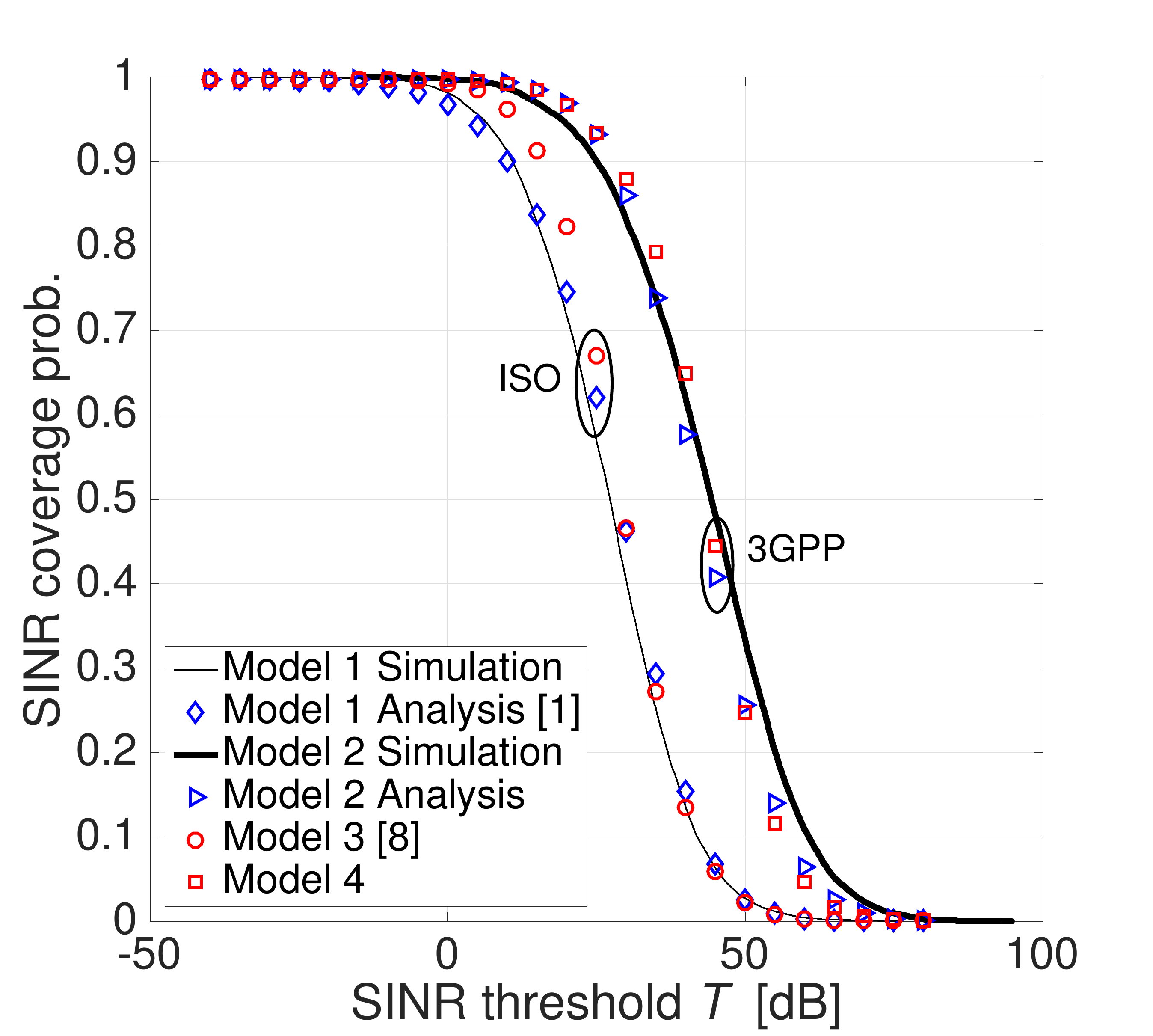}
\caption{For $\{n_{\text{TX}},n_{\text{RX}}\}=\{256, 64\}$.}
\label{comparison_256}
\end{subfigure}
\caption{Comparison between the $\SINR$ coverage probability expressions under Models 1 and 2 and their simplified expressions under Models 3 and 4. The simulated curves are obtained only under Models 1 and 2 without simplifying the channel-antenna configurations.
}
\label{global_comparison}
\end{figure*}

Finally, Fig.~\ref{global_comparison} illustrates the simplified $\SINR$ coverage probability expressions provided in Proposition~3 under Models~3 and 4 that are specified in Remark~5. As conjectured at the beginning of Sect.~\ref{Sect:simplified}, the simplified $\SINR$ coverage probability expressions become more accurate approximations for the $\3G$ element pattern than for the $\ISO$ element pattern. Precisely, the maximum difference between the simulated and the analytic $\SINR$ coverage probabilities are obtained as $7.7$\% with the $\3G$ element pattern and as $9.5$\% with the $\ISO$ element pattern in Fig.~\ref{comparison_256}.
This originates from both aligned and misaligned gains' identical tail behaviors that follow an exponential-logarithmic distribution.
These high-order behaviors are thus canceled out during the $\SINR$ calculation, and the first-order statistics thereby becomes dominant, from which the first-moment approximation used in the simplified $\SINR$ coverage expressions benefit.
On the contrary, with $\ISO$ element pattern, the aligned gain and misaligned gains have different tail behaviors as provided in Remarks~1 and 3, and the corresponding simplified $\SINR$ coverage probability expression therefore becomes less accurate. 

Moreover, the figure describes the benefit of the non-simplified $\SINR$ coverage probability expressions provided in Propositions~1 and 2 respectively under Models~1 and 2.
In contrast to the simplified expressions that are plausible only with the $\3G$ element pattern, the non-simplified $\SINR$ coverage probability expressions well approximate the simulated $\SINR$ coverage probabilities with both $\3G$ and $\ISO$ element patterns, so long as the number of antenna elements is sufficiently large, as seen by comparing Figs.~\ref{comparison_64} and~\ref{comparison_256}.
In addition, with a slight increase in complexity, these non-simplified SINR coverage probability expressions are more accurate than the simplified expressions, and so are appropriate for investigating ultra-reliable scenarios as considered in \cite{PetarURLLC:17,MehdiURLLC:18,UR2Cspaswin:17,giordani18}, which prefer to maximize accuracy rather than improving analytical tractability. It is also noted again that the simplified aligned and misaligned gains in Remark~5 are only applicable for the $\SINR$ calculation.
Thus, the non-simplified aligned and misaligned gains in Remarks~1-4 are still useful, for instance when deriving the mmWave interference distribution \cite{rebato16_interference} or calculating the mmWave signal-to-noise ratio (SNR) under a noise-limited regime \cite{rebato16_interference,fischione18}.

\section{Conclusions and Future Directions}
\label{conclusion_and_future_works}
In this study we have highlighted the impact of realistic mmWave channel behaviors and element patterns on the downlink $\SINR$ coverage probability in a large-scale mmWave network, via the NYU mmWave network simulator \cite{mezzavilla15} under the $\3G$ element pattern model \cite{antenna_3gpp}.
By introducing the aligned and misaligned gains, we have provided an analytical model that captures such realistic channel-antenna gain characteristics, thereby deriving the $\SINR$ coverage probability expressions. 

Especially for the $\3G$ element pattern, arguably the most practical antenna configuration, we proposed a further simplified $\SINR$ coverage probability expression.
This relies only on the exponentially distributed aligned and misaligned gains, which are known to be the simplest random variables for deriving the $\SINR$ coverage probability expressions. 

With a slight increase in complexity, we have also provided non-simplified $\SINR$ coverage probability expressions as well as the corresponding aligned and misaligned gain distributions.
These analytic expressions are versatile, and thus are expected to be exploited in more generic scenarios that particularly necessitate a higher accuracy, which could be an interesting topic for further research.
Furthermore, with the proposed analytic framework, an extension of this work could be to investigate other mmWave network settings such as different carrier frequencies, channel/antenna models, and an uplink scenario.
Besides, beyond the specific examples treated in the paper, our proposed methodology approach can be applied to study other cases.

\section*{Appendix I -- Proof of Proposition 1}
Consider the joint probability $\Pr \big( \SINR \geq T,  x_o\in \Phi_i\big)=\Pr \big( \SINR_i \geq T\big)$ in~\eqref{sinr_coverage_definition} when the typical UE associates with a BS in state $i\in \{L,N\}$. Applying the $\SINR$ definition in~\eqref{equation_sinr} to~\eqref{sinr_coverage_definition}, it is recast as follows.

\vspace{-10pt}\setcounter{equation}{37}
\footnotesize\begin{align}
\Pr \big( \SINR_i \geq T\big) &=\E_{r_{x_o}^i, I_i^L, I_i^N}\[\Pr \bigg( \frac{G_o^{(\ISO)}\ell^i(r_{x_o}^i)}{(I_i^L + I_i^N) + \sigma^2} \geq T \bigg)\] \label{jointProb}\\
&\hspace{-25pt}= \E_{r_{x_o}^i, I_i^L, I_i^N}\[  \Pr \left( G_o^{(\ISO)} \geq \frac{T (I_i^L + I_i^N + \sigma^2)}{ \ell^i(r_{x_o}^i)} \right)  \] \label{Eq:PfProp1_pre1}\\ 
&\hspace{-25pt}= \E_{r_{x_o}^i, I_i^L, I_i^N}\[  F_{G_o}^{(\ISO)}\(\frac{T (I_i^L + I_i^N + \sigma^2)}{ \ell^i(r_{x_o}^i)};\mu_o\)   \] \label{Eq:PfProp1_pre2}
\end{align}\normalsize
The last step is because the innermost probability in \eqref{Eq:PfProp1_pre1} corresponds to $G_o^{(\ISO)}$'s CCDF $F_{G_o}^{(\ISO)}(y;\mu_o)$ with $y$ that equals $T(I_i^L + I_i^N + \sigma^2)/\ell^i(r_{x_o}^i)$. 

Next, applying $F_{G_o}^{(\ISO)}(y;\mu_o)= \exp(-\mu_o y)$ in Remark~1 to \eqref{Eq:PfProp1_pre2}, we obtain

\vspace{-10pt}\footnotesize\begin{align}
\eqref{Eq:PfProp1_pre2}= \E_{r_{x_o}^i, I_i^L, I_i^N}\Bigg[ \exp&\( \frac{-\mu_o T  \sigma^2}{ \ell^i(r_{x_o}^i)}\) \nn \\
&\exp\( \frac{-\mu_o T I_i^L }{ \ell^i(r_{x_o}^i)}\) \exp\( \frac{-\mu_o T I_i^N}{ \ell^i(r_{x_o}^i)}\)  \Bigg]\\
= \E_{r_{x_o}^i}\Bigg[ e^{- \frac{\mu_o T  \sigma^2}{ \ell^i(r_{x_o}^i)}}& \E_{I_i^L}\Bigg[e^{ -\frac{\mu_o T I_i^L }{ \ell^i(r_{x_o}^i)}} \Bigg] \E_{I_i^N}\Bigg[e^{-\frac{\mu_o T I_i^N}{ \ell^i(r_{x_o}^i)}}\Bigg]  \Bigg]. \label{Eq:PfProp1_pre3}
\end{align}
\normalsize
The last step is firstly because $r_{x_o}^i$ is independent of $I_i^L$ and of $I_i^N$, according to Slyvnyak's theorem~\cite{HaenggiSG}. It is additionally because $I_i^L$ and of $I_i^N$ are mutually independent owing to the Markov property for the PPPs $\Phi_i^L$ and $\Phi_i^N$~\cite{HaenggiSG}. The innermost two expectation terms in~\eqref{Eq:PfProp1_pre3} can be represented using the Laplace transform $\mathcal{L}_X(s):=\E_X[e^{s X}]$. Then, the outermost expectation can be calculated using $r_{x_o}^i$'s PDF $f_{r_{x_o}^i}$ in~\eqref{Eq:DistPDF}, yielding

\vspace{-10pt}\footnotesize\begin{align}
\eqref{Eq:PfProp1_pre3}&= \int_0^{\infty} f_{r_{x_o}^i} \left(r\right) 
\exp \left(\frac{ -\mu_o T  \sigma^2 }{\ell^i(r)} \right)\mathcal{L}_{I_i^L}\left( \frac{\mu_o T}{ \ell^i(r)} \right) \mathcal{L}_{I_i^N}\left(\frac{\mu_o T}{\ell^i(r)}\right) \mathrm{d}r. \label{Eq:SIR_PfProp1}
\end{align}
\normalsize

Lastly, in what follows we expand $\mathcal{L}_{I_i^j}(s)$ with $s=\mu_o T/\ell^i(r)$ in \eqref{Eq:SIR_PfProp1}, i.e., the Laplace transform of the interference from the BSs in $\Phi_j$ for $j\in\{L,N\}$ when $x_o\in\Phi_i$. Following the interference expression in~\eqref{equation_sinr}, its Laplace transform is represented as follows

\vspace{-10pt}\footnotesize
\begin{align}
&\mathcal{L}_{I_i^j}(s) =  \E_{\Phi_{j},G_x}\left[\exp{\left(-s \sum_{x \in \Phi_j } G_x^{(\ISO)}  \ell^j(r_x) \right)}\right]\\
&\overset{(a)}{=} \E_{\Phi_j}\left[\prod_{x \in \Phi_j} \E_{G_x}\left[ \exp{\left(-s G_x^{(\ISO)}  \ell^j(r_x) \right)}\right]\right] \\
&\overset{(b)}{=} \exp \Bigg( -2 \pi \lambda_b \int_{\left( \frac{\beta_j r^{\alpha_i}}{\beta_i} \right)^\frac{1}{\alpha_j}}^{\infty} \left(1- \E_{G_x} \left[e^{-sG_x^{(\ISO)}  \ell^j(v)}\right]\right) v p_j(v) \mathrm{d}v \Bigg),
\label{Eq:Pf}
\end{align}\normalsize
step $(a)$ follows from the fact that $G_x^{(\ISO)}$ is independent of $\Phi_j$ and from i.i.d. $G_x^{(\ISO)}$'s.
Step $(b)$ comes from applying the probability generating functional (PGFL) of a HPPP~\cite{HaenggiSG}. Since the interfering BS locations and $G_x^{(\ISO)}$'s are independent, \eqref{Eq:Pf} is recast as follows

\vspace{-10pt}\footnotesize\begin{align}
\eqref{Eq:Pf} = \nonumber\\
\exp \Bigg( & -2 \pi \lambda_b \E_{G_x} \left[ \int_{\left( \frac{\beta_j r^{\alpha_i}}{\beta_i} \right)^\frac{1}{\alpha_j}}^{\infty} \left(1-  e^{-   \frac{\mu_o T G_x^{(\ISO)} \ell^j(v)}{\ell^i(r)} }\right) v p_j(v) \mathrm{d}v \right] \Bigg).
\end{align}\normalsize
The innermost expectation can be calculated using ${G_x}^{(\ISO)}$'s PDF $f_{G_x}^{(\ISO)}(y;a,b)$ in Remark~3. Combining this result with~\eqref{Eq:SIR_PfProp1} and~\eqref{sinr_coverage_definition} and applying the law of total probability completes the proof.~$\hfill\blacksquare$

\section*{Appendix II -- Proof of Proposition 2}
Replacing the exponentially distributed $G_o^{(\ISO)}$ by the $G_o^{(\3G)}$ in the joint probability calculation eq.~\eqref{jointProb}, we get
\footnotesize\begin{align}
\Pr \big( \SINR_i \geq T\big) &=\E_{r_{x_o}^i, I_i^L, I_i^N}\[\Pr \bigg( \frac{G_o^{(\3G)}\ell^i(r_{x_o}^i)}{(I_i^L + I_i^N) + \sigma^2} \geq T \bigg)\]\\
&\hspace{-25pt}= \E_{r_{x_o}^i, I_i^L, I_i^N}\[  F_{G_o}^{(\3G)}\(\frac{T (I_i^L + I_i^N + \sigma^2)}{ \ell^i(r_{x_o}^i)};b_o,p_o\)   \]. \label{Eq1:PfProp2}
\end{align}\normalsize
Similarly as before, the last step is because the innermost probability corresponds to $G_o^{(\3G)}$'s CCDF $F_{G_o}^{(\3G)}(y;b_o,p_o)$ with $y$ that equals $T(I_i^L + I_i^N + \sigma^2)/\ell^i(r_{x_o}^i)$. 

Next, applying $F_{G_o}^{(\3G)}(y;b_o,p_o)= \frac{\ln \left( 1-\left(1-p_o\right) e^{-b_o y} \right)}{\ln p_o}$ in Remark~2 to \eqref{Eq1:PfProp2}, we obtain

\vspace{-10pt}\footnotesize\begin{align}
\eqref{Eq1:PfProp2}= \E_{r_{x_o}^i, I_i^L, I_i^N}\bigg[ &\frac{1}{\ln (p_o)} \ln \bigg( 1 - (1 - p_o)\exp\bigg( \frac{-b_o T  \sigma^2}{ \ell^i(r_{x_o}^i)}\bigg) \nn\\
&\exp\bigg( \frac{-b_o T I_i^L }{ \ell^i(r_{x_o}^i)}\bigg) \exp\bigg( \frac{-b_o T I_i^N}{ \ell^i(r_{x_o}^i)}\bigg) \bigg)  \bigg]\label{Eq2:PfProp2}\\
\le \E_{r_{x_o}^i}\bigg[ \frac{1}{\ln(p_o)} & \ln \bigg( 1 - (1 - p_o) e^{- \frac{b_o T  \sigma^2}{ \ell^i(r_{x_o}^i)}} \nn \\
&\E_{I_i^L}\bigg[e^{ -\frac{b_o T I_i^L }{ \ell^i(r_{x_o}^i)}}  \bigg] \E_{I_i^N}\bigg[e^{-\frac{b_o T I_i^N}{ \ell^i(r_{x_o}^i)}}\bigg] \bigg) \bigg]. \label{Eq3:PfProp2}
\end{align}
\normalsize
The last step is firstly because $r_{x_o}^i$ is independent of $I_i^L$ and of $I_i^N$, according to Slyvnyak's theorem~\cite{HaenggiSG}.
Additionally because $I_i^L$ and of $I_i^N$ are mutually independent owing to the Markov property for the PPPs $\Phi_i^L$ and $\Phi_i^N$~\cite{HaenggiSG}.
However, differently from the proof of Proposition 1, here we used Jensen's inequality to derive an upper bound of~\eqref{Eq2:PfProp2}.
This permits to bring the expectations inside the logarithm, thanks to the fact that CCDF of $G_o^{(\3G)}$ is a concave function.
Then, the outermost expectation can be calculated using $r_{x_o}^i$'s PDF $f_{r_{x_o}^i}$ in~\eqref{Eq:DistPDF}, yielding

\vspace{-10pt}\footnotesize\begin{align}
\eqref{Eq3:PfProp2}&= \int_0^{\infty} \frac{f_{r_{x_o^i}} \left(r\right)}{\ln \left(p_o\right) }
\ln \bigg(1- \left(1-p_o \right)\nn\\
&\times \exp \left(\frac{ -b_o T r^{\alpha_i} \sigma^2 }{\beta_i} \right)
\mathcal{L}_{I_i^L}\left( \frac{b_o T}{ \ell^i(r) }\right) \mathcal{L}_{I_i^N}\left(\frac{b_o T}{\ell^i(r) }\right)\bigg)\mathrm{d}r. \label{Eq:SIR_PfProp2}
\end{align}
\normalsize

To conclude the proof, the Laplace transforms are used as in eq.~\eqref{Eq:Pf} of the proof of Proposition 1, where the interfering gain $G_x^{(\ISO)}$ is replaced with the respective $G_x^{(\3G)}$.~$\hfill\blacksquare$

\bibliographystyle{IEEEtran}
\bibliography{biblio}

\end{document}